\documentclass[aps,prb,reprint,superscriptaddress,twocolumn,showkeys,amsmath,amssymb,longbibliography,floatfix]{revtex4-2}
\usepackage{amsmath,amssymb,bbm,mathrsfs,bm,braket,color,graphicx,comment,amsfonts,dsfont}
\usepackage[colorlinks,linkcolor=blue,citecolor=blue,urlcolor=blue]{hyperref}
\usepackage[mathscr]{euscript}
\usepackage{physics}
\usepackage{xcolor}
\usepackage[normalem]{ulem}
\usepackage{bm}
\usepackage{orcidlink}
\usepackage{multirow}
\usepackage{microtype}

\input{macros.sty}
\usepackage[capitalise]{cleveref}
\usepackage{times}

\usepackage{pifont}

\begin{document}

\title{
Leveraging structural disorder to enhance topological phases 
}
\author{Laura Gómez Paz\orcidlink{0009-0001-0035-6598}}
\email{laura.gomez-paz@neel.cnrs.fr}
\affiliation{\small Universit\'e Grenoble Alpes, CNRS, Grenoble INP, Institut N\'eel, 38000 Grenoble, France}
\author{Peru d'Ornellas\orcidlink{0000-0002-2349-0044}
}
\affiliation{Donostia International Physics Center, P. Manuel de Lardizabal 4, 20018 Donostia-San Sebastian, Spain}
\author{Adolfo G. Grushin\orcidlink{0000-0001-7678-7100}}
\email{grushin@dipc.org}
\affiliation{Donostia International Physics Center, P. Manuel de Lardizabal 4, 20018 Donostia-San Sebastian, Spain}
\affiliation{IKERBASQUE, Basque Foundation for Science, Maria Diaz de Haro 3, 48013 Bilbao, Spain}
\affiliation{\small Universit\'e Grenoble Alpes, CNRS, Grenoble INP, Institut N\'eel, 38000 Grenoble, France}

\date{\today}

\begin{abstract}
On-site disorder can be leveraged to induce a transition from a trivial to a topological insulator. However it is unclear if structural disorder in the absence of on-site disorder can aid a similar transition and, if so, which kind of structural disorder is more favourable. We numerically show that structural disorder can enhance and sustain a topological phase up to strong disorder in two dimensions provided that one penalizes atomic sites from being close to one another. 
However, we find this effect is absent in three dimensions, where structural disorder appears generically detrimental to the phase.
In our calculations we include disorder that can scramble the global spin-reference frame, an overlooked type of disorder expected to exist in strongly disordered solids. This disorder fatally scrambles the information necessary for the spin-Bott and the spin-Chern marker to correctly diagnose a topological phase. 
By using the spectral localizer, a local marker directly defined using the time-reversal symmetry operator rather than a spin-projection, we show how one can circumvent this limitation, providing a basis-indifferent theory for calculating $\mathbb Z_2$ invariants.
Our work showcases that not all structural disorders are equally beneficial to topology, and highlights guiding principles to enhance and detect topological phases in both solid-state and metamaterial realizations.

\end{abstract}

\maketitle


\section{Introduction}

The presence of disorder in condensed matter systems is usually unavoidable, but its effect is not always detrimental. A remarkable example was the discovery that topological phases of matter can be induced from a topologically trivial state using on-site potential disorder~\cite{li_topological_2009,groth_theory_2009,Guo2010,Prodan2011,yamakage_z2_2012}. This is not a universal phenomenon since different kinds of disorder may or may not be favourable to stabilize topology~\cite{Song2012}.

Finding which disorder types destroy topological phases, and which favour them, requires calculating topological invariants in real space. Currently, a variety of such invariants exist, with varying regimes of validity, that allow us to characterise when non-trivial topology can be enhanced in disordered crystals 
\cite{li_topological_2009,groth_theory_2009,Guo2010,Prodan2011,Prodan2011b,Song2012,Ringel2012,Lv2013,fulga_statistical_2014,Ni2020,Silva2024,Chaou2025}, 
amorphous solids~\cite{agarwala_topological_2017,mansha_robust_2017,xiao_photonic_2017,mitchell_amorphous_2018,bourne_non-commutative_2018,poyhonen_amorphous_2018,minarelli_engineering_2019,chern_topological_2019,mano_application_2019,costa_toward_2019,marsal_topological_2020,mukati_topological_2020,sahlberg_topological_2020,ivaki_criticality_2020,agarwala_higher-order_2020,Grushin2020,Huang2020,wang_structural-disorder-induced_2021,focassio_structural_2021,Focassio2021,Mitchell2021,spring_amorphous_2021,wang_structural_2022,Ma2022,Peng2022,uria-alvarez_deep_2022,guzman_geometry_2022, spillage_2022,Cassella2022,Grushin2023,Cheng23,zhang2023anomalous,Manna2024,Li2024,marsal_obstructed_2022, uria2025}, 
quasicrystals~\cite{Kraus2012,Tran:2015cj,Bandres2016,Fuchs:2016hp,Huang2018,Huang2018,Fuchs:2018dd,Loring2019,varjas_topological_2019,Huang2019,Chen2019,He2019,Duncan2020,Fan2021,Zilberberg:21,Else2021,Hua2021,Jeon2022,Schirmann2024,Manna2024,rochecarrasco2025,caiger2026fractaltopologymajoranabound}, 
fractals~\cite{Marta2018,Manna2022,Manna2024}, or hyperbolic lattices~\cite{Urwyler2022,Liu_ChernHyperbolic_2022,Lenggenhager2022}.
There is a wide library of tools for calculating topological invariants, the integers that classify topological phases, directly in real space.
The Chern number, the integer ($\mathbb{Z}$) which classifies quantum Hall phases, has multiple expressions in real space~\cite{Kitaev20062,bianco11,marrazzo_locality_2017, Loring2010,LORING2015,loring2019guidebottindexlocalizer,DOrnellas2022,Hannukainen2022,Prodan2010,Prodan2011,Jezequel2022,jezequel2023modeshell,jezequel2025modeshell,Hannukainen2024,Bau2024,Favata2025}. 
In turn, real-space local markers for $\mathbb{Z}_2$ invariants are more challenging to define~\cite{prodan_robustness_2009,Akagi2017,Hannukainen2022,Gilardoni2022,favata_single-point_2023,Bau2024,Bau2024b,Huang2018,Huang2018b,Huang2019,Huang2020,Ni2020,wang_structural_2022,Ma2022,Cheng23}.
Some of them, like the spin-Bott index~\cite{Huang2018,Huang2018b} and the spin-local Chern marker~\cite{prodan_robustness_2009} require that a spin component is approximately preserved---essentially encoding a sensitivity to the basis in which one works.
While solutions to this have been proposed~\cite{Bau2024b}, other local markers do not naively suffer from this problem.
In particular, the spectral localizer index \cite{LORING2015,loring2019guidebottindexlocalizer,loring2019spectral, loring2020spectral, schulz-baldes_spectral_2021,doll2021skew, schulz2022invariants, cerjan_operator-based_2022, cerjan_local_2022,
schulz2023spectral,cerjan2023spectral,franca2024topological,
Cerjan2024crystal,Cerjan_2024_Tutorial,doll2024local, schulz2024topological,stoiber2024spectrallocalizerapproachstrong,Lee2025,wong2026} does not rely on defining a direction in spin space, so is manifestly basis independent.
Lastly, a numerically efficient alternative are scattering invariants~\cite{fulga_scattering_2012}, which one can show are equivalent to counting edge modes.

Often overlooked, a possible basis-dependence is especially important when working with amorphous matter. Ref.~\cite{Schirmann2025} pointed out that such systems can realize \textit{local-frame disorder}, in addition to connectivity, onsite and hopping disorder. 
Frame disorder describes the scrambling of the overall reference frame for the internal degrees of freedom at each site.
For example, because amorphous system lack an overall preferred direction, it is conceivable that the spin quantization axis can vary randomly between sites, a disorder we term \textit{spin-frame disorder}.
This effect is similar to adding Rashba spin-orbit coupling, where the spin-quantization axis ceases to be well defined~\cite{prodan_robustness_2009,Bau2024b}.

In this work we first highlight how considering spin-frame disorder exposes the limitations of the spin-Bott~\cite{Huang2018,Huang2018b,Huang2019,Huang2020,Ni2020,wang_structural_2022,Ma2022,Cheng23}, spin-Chern number~\cite{prodan_robustness_2009,Prodan2011,Gilardoni2022,favata_single-point_2023,Bau2024b} and the spectral localizer~\cite{LORING2015,loring2019guidebottindexlocalizer,doll2021skew,Setescak2025}, all widely used to identify the $\mathbb{Z}_2$ quantum-spin Hall topological phases. 
Specifically, the spin-Bott and Chern markers fail because locally scrambling the spin generically closes the spin gap in all directions.
In some realistic crystalline scenarios, spin-scrambling terms like Rashba spin-orbit coupling can be small enough to keep the gap open~\cite{Bau2024b}, but it is unlikely that in amorphous systems spin-frame disorder remains small.  Moreover, the spin-gap is always closed in three-dimensional $\mathbb{Z}_2$ topological insulators~\cite{Lin2024}.
Because the existence of a global spin gap is required for these operators to be well-defined, they thus fail to diagnose topology.
Ref.~\cite{Bau2024b} ingeniously circumvented this problem by defining a local $\mathbb{Z}_2$ marker via the time-reversal symmetry operator alone. 
It has the advantage of not relying on the spin-gap at the expense of introducing additional constraints to find projectors that are not related by time-reversal symmetry, which requires an additional optimization step.

Here we show how the spectral localizer, unlike other local markers, can be straightforwardly computed in any system regardless of the state of the spin gap. We find that the localizer explicitly depends on the time-reversal operator when computing $\mathbb Z_2$ invariants. We derive the exact method for computing the localizer for an arbitrary time-reversal operator, and show how this ensures that the resulting quantity is basis-invariant. This is particularly useful because current formulations of the spectral localizer typically assume a basis where time reversal is given by $i \sigma_y$ coupled with complex conjugation \cite{Loring2010,Cerjan_2024_Tutorial}.  
The formulation we present here allows the spectral localizer to remain well-defined and invariant for each spin-frame disorder realization and by extension, any effect that closes the spin gap. 

\begin{figure}[t] 
    \centering
    \includegraphics[width=\linewidth]{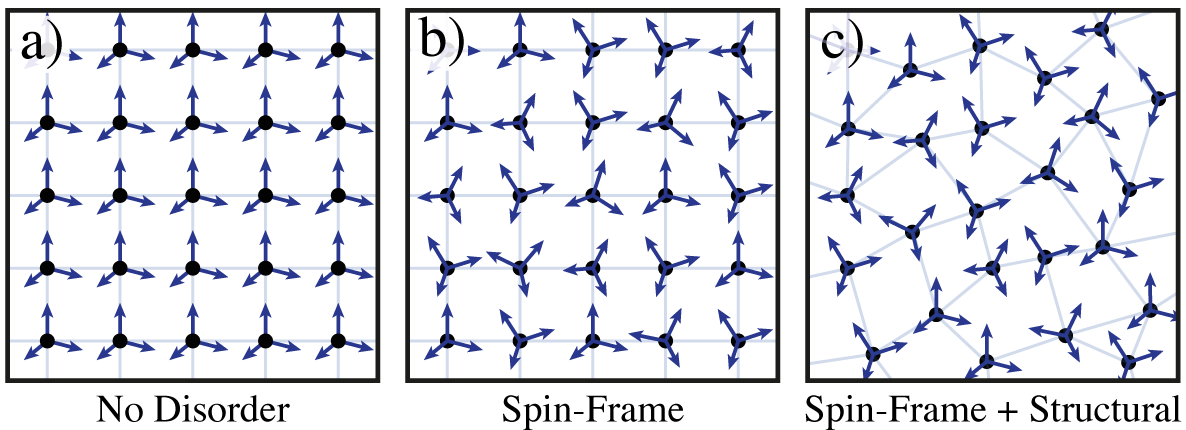}
    \caption{
    A comparison between different types of disorder. (a) In a crystal the spin quantization reference frame and coordination is equal across sites. (b) Spin-frame disorder, scrambles the spin-quantization reference frame between sites. (c) In amorphous systems, spin-frame disorder can coexist with structural, bond and on-site disorder. We study the effect of these disorders and point out limitations that they bring to known real-space methods to characterise topology.}
    \label{fig:intro_cartoon}
\end{figure}

We use this updated spectral localizer to comprehensively characterise two- and three-dimensional $\mathbb Z_2$-topological systems with onsite, hopping, structural and spin-frame disorder, see \cref{fig:intro_cartoon}.
We find that structural disorder can abruptly expand the phase space occupied by the non-trivial topological phase with respect to the crystalline limit.
This result expands existing results in crystals and non-crystalline amorphous systems where it is known that on-site potential disorder can induce a transition to the topological Anderson insulator phase~\cite{li_topological_2009,groth_theory_2009,Guo2010,Prodan2011,yamakage_z2_2012,Huang2018,Huang2018b}, while hopping disorder is typically detrimental to topology~\cite{Song2012}. 

Our study reveals that structural disorder involves the interplay of two distinct effects. Firstly, as a crystalline sample is disordered, the coordination number (i.e. the number of neighbours) can increase as sites move (on average) closer to one another, which can lead to an expanded topological phase space. 
This situation is more relevant to solids where metallic-type bonding allows for a disorder realisations to have a more densely packed structure compared to their crystalline counterparts, and less so to covalently bonded amorphous solids which typically preserve the number of neighbours~\cite{zallen_physics_1998,marsal_obstructed_2022}. Secondly, we find that fully random disorder can lead to a clustering, where the lattice effectively separates into groups of highly connected sites, adjacent groups weakly connected to one another. 

In order to separate the effect of coordination and clustering, we consider different mechanisms of structural disorder. We observe that introducing a repulsion between sites in the disorder limits this clustering but has no effect on the average coordination of the material. Thus, we find that while the increased average coordination has the effect of enhancing the topological regime, the clustering effect is detrimental to it. This interplay leads to a phase diagram with a strong dependence on the \textit{type} of structural disorder. We study both two- and three-dimensional systems, finding that the effect is more pronounced in two dimensions.

This work is structured as follows. In \cref{sec:model} we introduce the models for $\mathbb Z_2$ topological insulators that we consider. In \cref{sec:realspacetopo} we review the spectral localizer, the spin-local Chern marker and the spin-Bott index and compare their usefulness in the context of a disordered $\mathbb Z_2$ topological insulators. In \cref{sec:disorder} we study how spin-frame disorder affects the different markers and study how structural disorder changes the topological phase diagram compared to on-site Anderson disorder. Lastly we offer concluding remarks in Section \ref{sec:conclusions}. Technical details are left for the appendices. Appendix \ref{apx:lattices} describes how we construct systems that can interpolate between crystalline and amorphous point-sets, including the pair repulsion that distinguishes different types of structural disorder. Appendix \ref{app:kappa} and \ref{app:locsubtelties} include benchmarks on the spectral localizer's free parameter $\kappa$ and details on how the spectral localizer changes under arbitrary unitary rotations. Appendix \ref{app:markers} reviews the Bott index and Chern marker for class A and derives their equivalence. Lastly, Appendix \ref{apx:finitesizescaling} includes checks on finite size convergence.

\section{Models for \texorpdfstring{$\mathbb Z_2$}{Z2} Topological Insulators} \label{sec:model}

In both the two- and three-dimensional case, we consider a Hamiltonian defined on a (potentially amorphous) lattice, represented this as a set of points at positions $\textbf r_{j}$. Each site hosts two orbitals and a spin-$1/2$ degree of freedom---leading to four states per real-space site. The Hamiltonian is parametrised as
\begin{equation}
\label{eq:hamgen}
    H = \sum_{\langle jk\rangle} f_{jk} \ketbra{\textbf r_j}{\textbf r_k}
    \otimes  t_{jk}
    +\sum_{j} \ketbra{\textbf r_j}{\textbf r_j}
    \otimes \epsilon_{j},
\end{equation}
where $f_{jk}$ is a scalar that controls the strength of the hopping, $t_{jk}$ represents a hopping matrix and $\epsilon_j$ is a matrix acting on the internal degrees of freedom at site $j$. We parametrise $t_{jk}$ and $f_{jk}$ in terms of the displacement vector $\bm \delta_{jk} = \textbf r_j - \textbf r_k$. We choose the hopping strength to depend exponentially on the magnitude of the separation, $\delta_{jk}=|\bm \delta_{jk}|$, as~\cite{agarwala_topological_2017}
\begin{align} \label{eqn:coupling_strength}
    f_{jk} =  e^{-(\delta_{jk} - r_0)/r_0 }\Theta(R-\delta_{jk}),
\end{align}
where we choose a cut-off distance of $R=1.3r_0$ in units of $r_0$, the effective average separation $r_0 = {L}/{\sqrt[d] N}$,  with $L$ the linear system size, $N$ the number of sites and $d$ the dimensionality of the system. 

In both two and three dimensions, the hopping term may be written almost identically, in terms of the normalised displacement from one site to its neighbour, $t_{jk} = t(\hat {\bm \delta}_{jk})$, with $\hat {\bm \delta}_{jk} = \bm \delta_{jk}/ \delta_{jk}$. The hopping is term is given by
\begin{align}
    t_{jk} = \frac{A}2 \left [
        \tau_z \otimes s_0 - i \tau_{x} \otimes (\hat{\bm \delta}_{jk} \cdot \bm s)
        \right ].
\end{align}
Here, the Pauli matrices $\boldsymbol{\tau}=(\tau_x,\tau_y,\tau_z)$ and $\bm s=(s_x,s_y,s_z)$ act on orbital and spin degrees of freedom respectively. In two dimensions, the displacement $\bm \delta$ is a two element vector, and thus we use only $\bm s = (s_x, s_y)$ in the hopping, whereas in three dimensions we use the full three-element $\bm s$. 
The identity matrices in these subspaces are $\tau_0$ and $s_0$. The angle dependence of the Hamiltonian is given by the Slater-Koster parametrization between orbitals located at neighbouring sites~\cite{slater_simplified_1954}.
The onsite energy is parametrised as
\begin{align}\label{eqn:ham_onsite}
    \epsilon_{j} &=
    M \tau_z \otimes s_0 + W_j  \tau_0 \otimes  s_0,
\end{align}
where $M$ is an orbital asymmetry and disorder at each site $j$ is drawn from a uniform distribution $W_j \in [-W/2,W/2]$.

In two dimensions, this Hamiltonian is equivalent to an analytic continuation of the Bernevig-Hughes-Zhang quantum-spin Hall model (BHZ)~\cite{roy_phase_2025,bernevig_quantum_2006}, which is well-defined on an amorphous point-set~\cite{agarwala_topological_2017}, and reduces to the conventional model on the square lattice \footnote{Strictly, the Hamiltonian written here is a basis rotation away from the canonical form that the BHZ model is usually written in.}. Thus, the model has an invariant $\nu_{\mathrm{2D}} \in \mathbb Z_2$ defined as 0 in the trivial phase and 1 in the topological phase.

On a crystalline square lattice and with no on-site disorder ($W_j=0~\forall j$) this model realizes a topological quantum-spin Hall phase in symmetry class DIII at half filling with $\nu_{\mathrm{2D}}=1$ index when $0<|M|<4$ and a trivial insulator phase with $\nu_{\mathrm{2D}}=0$ when $|M|>4$. The topological phase diagram is symmetric with respect to $M=0$~\cite{Yamakage2013}, so we will focus our study on $M<0$. Once $W_j\neq0$, particle-hole symmetry is lost and the model falls into class AII.

In three dimensions, our model realizes an analytic continuation of the three-dimensional topological insulator \cite{hasan2010,Qi_Zhang2011,Fu2007,Roy2009}, which also has a $\mathbb Z_2$ invariant, $\nu_{\mathrm {3D}}$.
On a crystalline cubic lattice with no on-site disorder ($W_j=0$), this model realizes a three-dimensional $\mathbb{Z}_2$ topological insulator phase in symmetry class DIII at half filling with $\nu_{\mathrm {3D}}=1$ when $1<|M|<3$, and a trivial and weak insulator phases with $\nu_{\mathrm {3D}}=0$ when $|M|>3$ and $0<|M|<1$, respectively. The topological phase diagram is symmetric with respect to $M=0$~\cite{Yamakage2013}, so we will focus our study on $M>0$. As in the two-dimensional case, adding disorder ($W_j\neq 0$) destroys particle-hole symmetry, putting the model into class AII.

\section{Real-space topology \label{sec:realspacetopo}}

The topological invariants governing the spin Hall effect were initially proposed in a crystalline context \cite{kane_z2_2005,sheng_quantum_2006,fu_time-reversal_2006}. However, the phase itself is highly robust against real-space disorder which destroys our ability to construct a momentum-space invariant. Consequently, a number of different quantities have been proposed to diagnose $\mathbb Z_2$ topological insulators in real space. The most commonly used are the spectral localizer \cite{LORING2015}, the local Chern marker \cite{bianco11} and the associated spin-Bott index \cite{Huang2018,Huang:2018gu}. Our aim here is to compare the usefulness of these markers in a generically disordered spin-Hall material in which neither momentum space, nor any local degree of freedom such as spin is a guaranteed good quantum number.

\subsection{Spectral localizer}

The construction of the spectral localizer operator, introduced in Ref.~\cite{LORING2015} and reviewed in Ref.~\cite{Cerjan_2024_Tutorial}, is designed to provide information about the localization of the Hamiltonian in both energy and position so that its spectral properties encode the topological phase. 
The general form of this operator for an arbitrary number of physical dimensions $d$ reads as
\begin{equation}
    \mathcal L_{(\mathbf{x},E)} = \kappa \boldsymbol{\Gamma} \otimes (\mathbf{X} - \mathbf{x}  \1)  + \Gamma_{d+1} \otimes (H - E\1),
    \label{eq:general_localizer}
\end{equation}
where $\mathbf{x}=(x_1,x_2,\cdots,x_{d})$ and $E$ center the spectrum at a specific position and energy respectively, and $\kappa \approx E_{\text{gap}}/L$ is a dimensionful scaling coefficient used to ensure unit consistency. There are $d+1$ $\Gamma_\mu = (\boldsymbol{\Gamma},\Gamma_{d+1})$ matrices that form a Clifford representation large enough to satisfy the last expression.
The choice of parameter \texorpdfstring{$\kappa$}{k} is crucial, as it balances the spectral weight of the Hamiltonian and position operators: a correct value of \texorpdfstring{$\kappa$}{k} would allow the localizer to see changes in both $H - E$ and $\mathbf{X} - \mathbf{x}$. A detailed explanation of the procedure for choosing this coefficient is given in Appendix~\ref{app:kappa}, and discussed extensively in Ref.~\cite{Cerjan_2024_Tutorial}. 

In two dimensions, we choose the Pauli matrices $\{\sigma_x, \sigma_y, \sigma_z\}$ as our representation  $\Gamma_n$ of the Clifford algebra, such that the spectral localizer is
\begin{align}
    \begin{aligned}
\label{eqn:2D_localizer}
    \mathcal L_{(\mathbf{x},E)} =& \kappa\left [\sigma_x \otimes (X-x\1) +
      \sigma_y \otimes(Y-y\1)\right ] \\
      &+  \sigma_z \otimes (H-E\1).
    \end{aligned}
\end{align} 
From now on, the position and energy shifts $(x,y, E)$
will be absorbed into the definitions of the corresponding operators, $X$, $Y$ and $H$.

 In phases with a $\mathbb Z$ invariant  such as the integer quantum Hall effect, one can prove that the signature of this operator, $\text{sig}[\mathcal L]$, defined as the half-difference between the number of positive and negative eigenvalues, determines whether the Hamiltonian can be continuously transformed to commute with the position operators without closing the band gap~\cite{Cerjan_2024_Tutorial}. This is crucial for determining the system's topology because the occupied wavefunctions of strong topological phases in dimensions $d>1$ cannot be exponentially localized~\cite{DJThouless_1984,Soluyanov2011,altland_fragility_2024}. When $\text{sig}[\mathcal L] = 0$, the system can be adiabatically connected to an atomic insulator where $H$ and $X_j$ commute without closing the gap, implying the system is in a trivial phase. A closing of the gap in the spectrum of the localizer along with a change in the signature, signals a transition to a different topological phase, with an invariant given by the signature of $\mathcal L$.

The Hamiltonian in Eq. \eqref{eq:hamgen} is time-reversal symmetric, with $\mathcal T^2 = -1$, placing the system in class AII in both two and three dimensions. We can then determine constructively the symmetry class of the spectral localizer matrix itself, this is done in Appendix \ref{app:locsubtelties}. We find that the localizer for the two-dimensional spin hall case is in Class D---with particle-hole symmetry only---such that eigenvalues come in pairs, and $\text{sig}[\mathcal L] = 0$. Therefore, we need to use a different object to diagnose topology. 
For the two-dimensional case we use the Pfaffian~\cite{Cerjan_2024_Tutorial}.
At the closings of the localizer gap (hereafter denoted as the local gap), the Pfaffian of $\mathcal L$ might change sign, in which case there is a topological phase transition.

Since the Pfaffian is only well-defined for an antisymmetric matrix, the localizer $\mathcal L$ must first be rotated to a basis in which it is antisymmetric and pure imaginary
\begin{align}
    \label{eqn:rot_localizer_main}
    \mathcal L_C &= Q^\dag \mathcal L Q,
\end{align}
where $Q \in SU(n)$ is matrix responsible for the rotation. Finding the appropriate matrix is not trivial for generic systems, since it depends explicitly on the exact formulation of the time-reversal operator. In Appendix \ref{app:locsubtelties} we construct an explicit method for finding the correct rotation for an arbitrary system, where we find that $Q$ is obtained by taking a Takagi factorisation \cite{takagi_algebraic_1924} of the time-reversal operator. Once this transformation has been found, the $\mathbb Z_2$-invariant for a two-dimensional system is $\nu_{\textup{2D}} \in \{0,1\}$ with
\begin{align}
\label{eq:index_loc}
(-1)^{\nu_{\textup{2D}}} &= \operatorname{sign}(\pf{i\mathcal L_C})  \in \mathbb Z_2.
\end{align}

For the three dimensional case, we choose 
$\Gamma_i= \{\sigma_y \otimes \sigma_x, \sigma_y \otimes \sigma_y, \sigma_y \otimes \sigma_z,\sigma_x \otimes \sigma_0\}$ with $i=1,\cdots,4$ as our representation of the Clifford algebra, such that the spectral localizer is expressed as
\begin{align}
    \label{eq:3D_localizer}
    \begin{aligned}
        \mathcal{L}_{(\mathbf{x},E)} =& \kappa\Big[\Gamma_1 \otimes (X-x\1) + \Gamma_2 \otimes (Y-y\1) \\ 
        &+ \Gamma_3 \otimes (Z-z\1)+\Gamma_4 \otimes (H-E\1) \Big] .
    \end{aligned}
\end{align}
This localizer matrix falls into Class BDI---with chiral, time reversal and particle hole symmetry. The calculation of the topological invariant in this case can be reduced to the sign of the determinant of the off-diagonal block of the localizer, arriving at $\nu_{\textup{3D}} \in \{0,1\}$ with
\begin{align}
    \label{eq:signdet}
(-1)^{\nu_{\textup{3D}}}  = \operatorname{sign}\big(\det\big(\sigma_0 \otimes H - i \kappa \big[  \boldsymbol{\sigma} \otimes \mathbf{X} \big]\big)\big) \in \mathbb Z_2.
\end{align}
Here $\boldsymbol{\sigma}$ is a vector of three Pauli matrices and $\mathbf{X}=(X,Y,Z)$, and we have absorbed the shifts in position and energy into the definitions of $H$ and $\mathbf{X}$.

\subsection{Bott and spin-Bott Index}

In two dimensions, one popular route towards a local invariant has been the construction of the Bott index~\cite{Loring2010}, and associated spin-Bott index~\cite{Huang2018,Huang2018b}. In the limit of no spin mixing, a $\mathbb Z_2$ spin-Hall insulator can be viewed as two copies of a conventional Chern insulator (class A in two dimensions), one populated entirely with spin-up and the other populated with spin-down electrons. Here the time-reversal symmetry enforces that both species have opposite Chern number---however in all other respects the two species may be treated as completely separable. This occurs when the Rashba spin-orbit coupling term, which mixes spin components, is set to zero, and when there is no spin-frame disorder. 

In this limit, constructing a real-space topological index is straightforward, since one can effectively use the local markers that have already been shown to be effective for Chern insulators \cite{Kitaev20062, bianco11}. Furthermore, after adding weak spin-mixing, this procedure still gives a reasonable local marker for the $\mathbb Z_2$ invariant \cite{Huang2018,Bau2024b}. Thus, let us briefly restate the form of the Bott Index, discussing its efficient numerical implementation, and then we shall explain the procedure for applying these quantities to a spinful system. 

The Bott index is defined for a system with periodic boundaries, where we construct the matrices $U = Pe^{-i\delta X}P$ and $V = Pe^{-i\delta Y}P$, where $P$ is the projector onto a gapped subset of the spectrum, generally the occupied states below the Fermi level, $X$ and $Y$ are the position operators, and $\delta = \frac{2\pi}L$. The Bott index is given by
\begin{align}
    \mathcal B(P) = \frac{L^2}{2\pi}\im \Tr \log \left [ UVU^\dag V^\dag + Q\right],
\end{align}
with $Q = \1 - P$. Since a matrix logarithm is generally an expensive computation, we may expand the expression in $\delta \ll 1$, a good approximation for large system size, to arrive at the following expression \footnote{This expression is similar to the Chern marker \cite{bianco11}, however is well-posed even in periodic boundaries, since the commutator $[X,P]$ is well-defined even if $X$ itself is not. See Appendix \ref{app:markers} for elaboration on this subtlety.},
\begin{align} \label{eqn:chern_bott}
    \mathcal B(P) =  2 \pi i \Tr \left ( P [X,P]_{\textup{PB}} [Y,P]_{\textup{PB}} \right ) + h.c. + \mathcal O(\delta^3),
\end{align}
where the \textit{periodic commutators}, $[\cdot ,\cdot ]_{\textup{PB}}$, are defined as
\begin{align}
    [X,P]_{\textup{PB}} = \sum_{jk} \Delta_{jk}^x P_{jk} \ketbra{\textbf r_j}{\textbf r_k},
\end{align}
with a similar expression for $Y$. Here $\bm \Delta_{jk}$ is defined as the shortest displacement between sites $\textbf r_j$ and $\textbf r_k$ \emph{in periodic boundaries}. Note that in open boundaries, these are exactly equal to a conventional commutator. For a full derivation of this quantity, see Appendix \ref{app:markers}. 

In order to apply the spin-Bott index to the spin-Hall case, we consider the projector onto our chosen (Kramers degenerate) band $P$. Our aim is to decompose $P$ into a spin-up and spin-down component, each of which may then be treated separately. Thus, we construct the projected spin operator $\tilde s^z = Ps^zP$. By diagonalising $\tilde s^z$, we can construct the two spin-resolved projectors onto the spin-up and down subspaces, $P_\uparrow$ and $P_\downarrow$ as the projector onto the positive and negative eigenspaces of $\tilde s^z$. Finally, the spin-Bott index is calculated according to
\begin{align} \label{eqn:spin-bott-index}
    \mathcal B_{\textup{spin}}(P) = \frac 12 
    \left [
    \mathcal B(P_{\uparrow}) - \mathcal B(P_{\downarrow})
    \right ].
\end{align}
Thus, the construction of the spin-Bott index is essentially analogous to the spin Chern number, with the Bott index replacing the momentum space Chern number. As with the spin Bott index, the validity of the spin Chern number depends on the spin gap---the smallest magnitude eigenvalue of $\tilde s^z$---remaining open \cite{prodan_robustness_2009}.

\section{Disorder \label{sec:disorder}}
Amorphous materials are expected to disorder the magnitude of on-site potentials (which we refer to as Anderson disorder), the hopping amplitudes (hopping or bond disorder), as well as disordering the connectivity of the lattice structure itself (structural disorder)~\cite{zallen_physics_1998,weaire_electronic_1971}. Historically, the most well-understood form is Anderson disorder~\cite{Anderson1958}, where a random on-site potential is applied across the system in real space---parametrised with $W_j$ in \cref{eqn:ham_onsite}. In quantum spin-Hall Hamiltonians, Anderson disorder can drive the system from a trivial phase into a topological one by inverting the gap~\cite{li_topological_2009,groth_theory_2009,Guo2010,Prodan2011,Prodan2011b,Song2012}. This topological Anderson phase can be smoothly connected to the crystalline phase without closing of the mobility gap and is, in this sense, the same quantum-spin Hall phase~\cite{Prodan2011}.

Because the hopping and the connectivity can be disordered, Anderson effects alone are insufficient for describing the interplay between topology and disorder in amorphous matter, where disorder can take many different forms. 
In this work we follow up on Ref.~\cite{Schirmann2025} and also consider \textit{frame disorder}, where the local degrees of freedom at each site are scrambled. To have a full understanding of how topological phases can emerge in amorphous matter, it is useful to separate each of these different manifestations of disorder, and understand the effect of each on the system's topological phase diagram.

\subsection{Spin-frame disorder}

\begin{figure}[t] 
    \centering
    \includegraphics[width=0.97\linewidth]{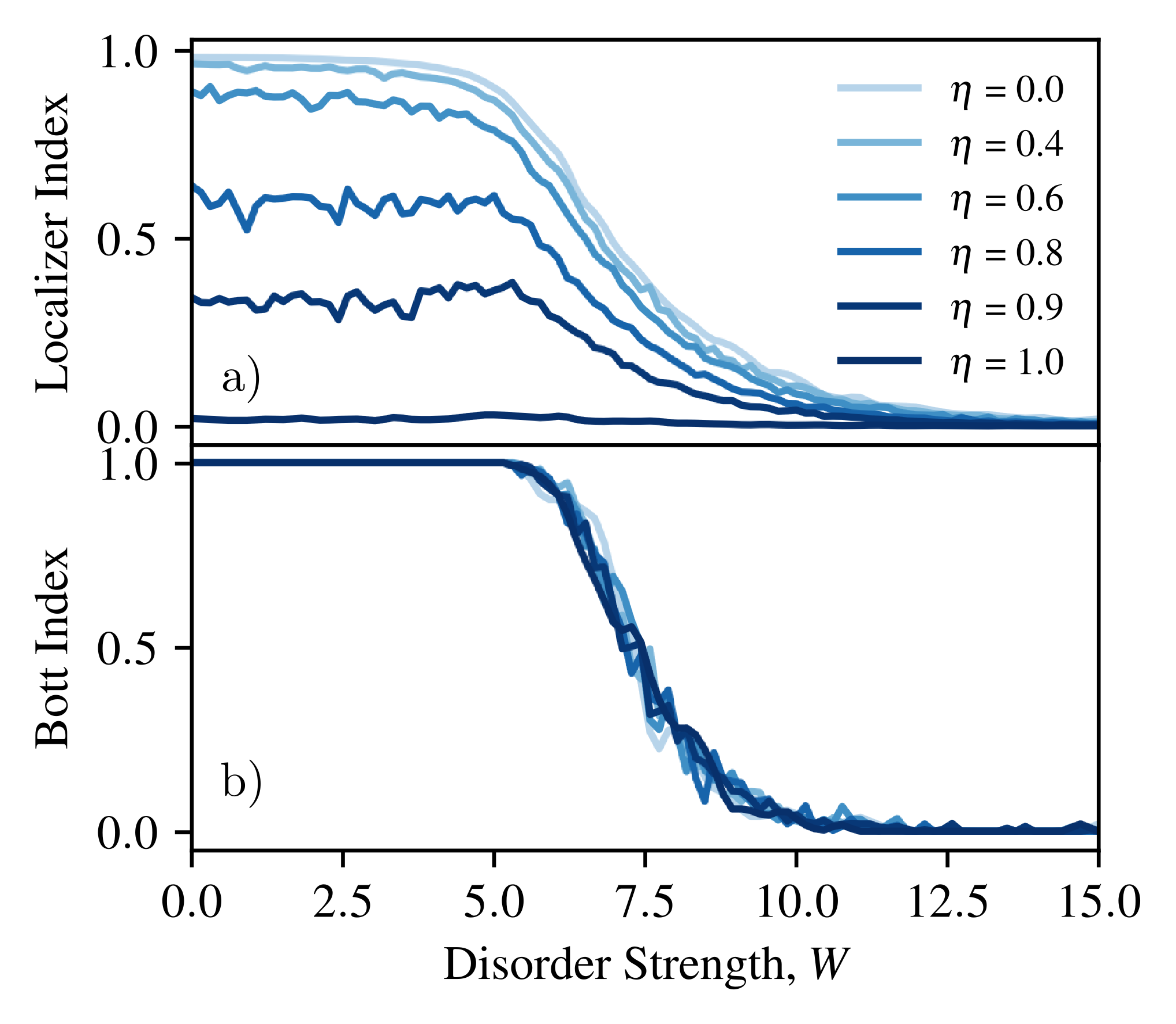}
    \caption{
    Failure of the spin-Bott index in the presence of spin-frame disorder. (a) As the fraction of sites with spin-frame disorder is increased from $\eta= 0$ to $\eta =1$ the spin-Bott index from \cref{eqn:spin-bott-index} continuously drops to zero. At $\eta=1$ if fails to diagnose the topological transition. The spin-Bott index requires a well defined spin gap, which is generically closed by spin-frame disorder. (b) The localizer index from \cref{eq:index_loc} (with $\kappa = 1$) is independent of $\eta$ and correctly captures the topological phase transition as a function of disorder strength $W$. }
    \label{fig:spin_frame_bhz}
\end{figure}

When a material is isotropic in the bulk on average, one can lose any notion of a special direction in both real- and spin-space, leading to a type of disorder called \textit{spin-frame disorder}. To model this, we randomly rotate the reference frame of local degrees of freedom at each site, thereby losing any global notion of spin-alignment, which inhibits our ability to calculate topological invariants that depends anisotropically on spin. Our goal is to show how local markers in class AII are sensitive to the orientation of the spin, which is scrambled by this frame disorder. For example, the spin-Bott index and spin-Chern marker assume a direction of spin-projection that is conserved, which is no longer true in the presence of frame disorder. The Localizer index does not assume such conservation, but it directly relies on identifying the time-reversal operator, which is basis dependent, see Appendix \ref{app:locsubtelties}. Frame disorder locally scrambles the basis of the Hamiltonian, acting as a unitary rotation on the time-reversal operator, which must be accounted for when calculating the spectral localizer index.

To analyse how frame disorder affects local markers we start by defining it precisely through a local unitary transformation applied to the Hamiltonian. As a simple example of a local unitary transformation that rotates the spin at each site, we consider the effect of a Hadamard gate, defined as
\begin{align}
    \mathsf H = \frac{1}{\sqrt 2}\begin{pmatrix}
        1 & 1 \\ 
        1 & -1
    \end{pmatrix}.
\end{align}
We define a random variable per site $a_{j}\in \{0,1\}$ with uniform probability $\eta$ that decides whether that site undergoes a Hadamard transformation. Thus, we may construct a global random rotation operator $U$, characterised by this random variable, according to 
\begin{align}
\label{eq:UHad}
    U = \sum_{j}\ketbra{\textbf r_j}{\textbf r_j} \otimes \mathsf H^{a_{j}}
\end{align}
and transform the system according to $ H \rightarrow U H U^\dag$.
If our Hamiltonian is time-reversal symmetric in a basis where this symmetry is represented by the operator $\mathcal{T}= is_y\mathcal K$, frame disorder will modify this operator to
\begin{align}
    \hat {\mathcal{T}} \rightarrow \hat {\mathcal{T}}' &
    = w \mathcal K,\\
    \textup{with } w &= U is_y U^T.
\end{align}
In the case of the Hadamard operator, which anti-commutes with $s_y$, this may be explicitly calculated as
\begin{align}
    w = \sum_j \ketbra {\textbf r_j}{\textbf r_j} \otimes (-1)^{a_j} i s_y.
\end{align}
This transformation must be accounted for when we construct the spectral localizer, since any change to the time-reversal operator must be taken into consideration when we rotate the localizer into a pure-imaginary basis, such that the Pfaffian is well-defined. An explicit derivation for the $\mathbb Z_2$ invariant in any basis---where the time-reversal operator can take any form---is given in Appendix \ref{app:locsubtelties}. Let us summarise the result here: the matrix $Q$ which brings the localizer to an antisymmetric form---given by $Q = \1 + i\sigma^y \otimes s^y$ for a conventional time-reversal-symmetric-invariant system with $\mathcal T = i s^y \mathcal K$--- must be modified by the same spatially-dependent rotation operator \cref{eq:UHad}, according to
\begin{align}
    Q \rightarrow U Q U^T.
\end{align}
To study the effects of spin-frame disorder we construct the two-dimensional BHZ model given in \cref{sec:model} on the square lattice with parameters $A=1$, $M = -2$, placing us in the center of the topological region of the phase diagram. 
We tune both the strength of Anderson disorder $W$, and the proportion of rotated sites $\eta$ across the full range $[0,1]$. Our expectation is that as $W$ is increased, the system should eventually be driven from a topological phase to a trivial phase~\cite{li_topological_2009,groth_theory_2009,Prodan2011}, which happens around $W \sim 7.5$. Spin frame disorder, however, should have no effect since it acts effectively as a basis rotation. Each time, we compute both the Bott index and spectral localizer. For each data point we average the invariant over 50 realisations of spin-frame and Anderson disorder. 

The results are shown in \cref{fig:spin_frame_bhz}. Considering \cref{fig:spin_frame_bhz} (a), we see that as spin-frame disorder is increased---by increasing the probability of rotating a local basis, $\eta$---the Bott index decreases and eventually vanishes across the whole phase diagram. This reflects the fact that in the case of high spin-frame disorder, the spin gap, upon which the index depends, closes, and so the spin Bott index is effectively ill-defined. On the other hand, in \cref{fig:spin_frame_bhz} (b), we see that the results obtained using the spectral localizer are indifferent to spin frame disorder.

\subsection{Structural disorder}
\begin{figure}
    \centering
    \includegraphics[width=0.97\linewidth]{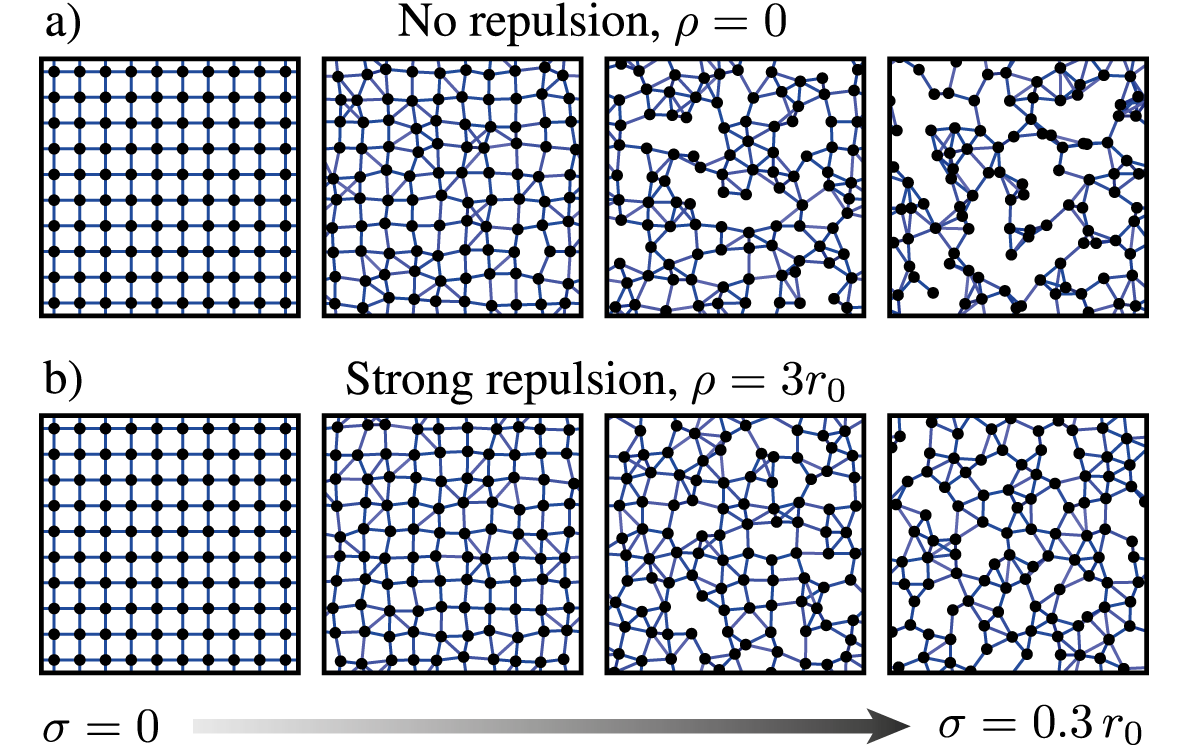}
    \caption{Two examples of progressively disordering a system starting from a square lattice on the unit square. At each point, bonds are drawn between pairs of sites that are closer than $1.3$ times the original lattice spacing $r_0$.  
    (a) With the repulsion parameter, $\rho$, set to zero, each point is moved to a new position sampled from a Gaussian distribution centred on the original position of the vertex with standard deviation $\sigma$. As $\sigma$ is increased, this tunes smoothly between the square lattice and a set of uniformly distributed vertices. 
    (b) With $\rho$ set to $0.3$, the probability of moving a vertex close to any other existing vertices is suppressed. Thus, as the system is progressively disordered by increasing $\sigma$, vertices move to a `blue noise' configuration, in which they are randomly arranged but evenly spaced from one another.
    }
    \label{fig:lattices}
\end{figure}

We now turn our attention to understanding the role of structural disorder. Here, we wish to tune smoothly between the square or cubic lattices and an amorphous lattice, tracking how the properties of the physical system change. Similar questions were discussed e.g.~in Ref.~\cite{wang_structural_2022}, where structural disorder combined with on-site Anderson disorder resulted in a topological phase induced by amorphization. However, it remains unclear whether structural disorder can drive the transition without relying on on-site disorder, and if it can enlarge the phase space occupied by the topological phase. Ref.~\cite{Ma2022} showed that hyperuniform structural disorder which kept the number of neighbours of each atom fixed is detrimental to topology. Here we show that structural disorder is not always detrimental when you allow the number of neighbours to fluctuate, and that this effect  depends on dimensionality. This serves to illustrate that the consequences of structural disorder are not universal; depending on the form of disorder used we can arrive at a substantially different phase diagram.

To this end we define a process for progressively disordering  a square lattice. This is done by introducing a random shift to each vertex in the system, sampled from a two-dimensional Gaussian distribution with standard deviation $\sigma$. As $\sigma$ is tuned from zero to a value comparable to the lattice spacing, the points are smoothly interpolated between a square grid and a completely random set of points. This is shown in \cref{fig:lattices} (a). Once the disordered point-set has been created, a network is constructed by adding bonds between all pairs of vertices that are closer than a threshold, which here is chosen to be $1.3r_0$---where $r_0$ is the lattice spacing of the original square lattice---giving the cut-off in \cref{eqn:coupling_strength}. Close to the crystallization temperature, $\sigma$ can be thought of as determined by the distance that atoms can move thanks to the thermal energy provided by the quenching protocol that forms the amorphous solid, such that $\sigma \propto k_B T$,~\cite{wang_structural_2022}.

Note that in the limit of high disorder ($\sigma \approx r_0$), this procedure creates highly unphysical arrangements of atomic positions, since randomly sampled points have a tendency to form clusters with extremely small spacing between vertices. In contrast, covalently bonded amorphous materials have strong local correlations due to chemical constraints~\cite{weaire_electronic_1971,zallen_physics_1998,marsal_topological_2020,corbae2023,Ciocys2023}. Thus, such random disorder, although useful as a benchmark for the effects of disorder, have limited physical applicability when considering real materials. 

To address this, we modify the probability distribution that is sampled each time a vertex is moved, adding an inter-site repulsion parametrised by $\rho$. As $\rho$ is increased, the probability of placing a site close to another existing site is reduced. This effectively sets a length scale with which vertices can `see' one another. This probability distribution is constructed in analogy to statistical mechanics, where we construct an effective energy associated with a configuration of points, and the probability distribution is given by the partition function derived from this energy. The Gaussian disorder ($\sigma$) is introduced as a quadratic potential well centred on the original position of each vertex, and the repulsion ($\rho$) appears as an interaction energy. The strengths $\sigma$ and $\rho$ therefore model two competing phenomena: the thermal energy and the tendency to form a well defined local chemical order. Such possible competition has been previously overlooked in the search for larger topological phase diagrams and is behind the richness in phenomenology that we report below.

Using the parameter $\rho$ we are able to tune between uniformly sampled points and a `blue noise'-like distribution \cite{mitchell_generating_1987,mitchell_spectrally_2991,fattal_blue_2011} in which points are random but approximately uniformly spaced apart, shown in \cref{fig:lattices} (b). The procedure used is described in detail in Appendix~\ref{apx:lattices}, along with a study of the average connectivity, bond lengths and structure factors obtained using this method.

In addition to modifying the connectivity of the lattice, we include a bond disorder implied by the lattice spacing, in which the coupling strength depends exponentially on the length of a bond, according to \cref{eqn:coupling_strength}. 
We recall that in the crystalline case, bond disorder is typically detrimental to the topological Anderson phase~\cite{Song2012}. To better distinguish a potential enhancement of the topological phase due to structural disorder, we do not add on-site Anderson disorder ($W_j=0, \forall j$), which is known to favor topology~\cite{li_topological_2009,groth_theory_2009,Guo2010,Prodan2011,yamakage_z2_2012}.

We compare the effect of Anderson disorder against structural disorder in both two and three dimensions, where we calculate the phase diagram as a function of both the internal parameter $M$ (setting $A = 1$) and the respective disorder strength. In the case of Anderson disorder---which serves as a benchmark---this is the potential strength $W$ in \cref{eqn:ham_onsite}. In the case of structural disorder we tune the standard deviation $\sigma$ determining the length-scale of the (Gaussian) probability distribution for the displacement of each vertex in the lattice. We consider three cases, one where $\rho = 0$, such that there is no inter-site repulsion, one where $\rho = 3r_0$ and the last with stronger inter-site repulsion $\rho = 9r_0$---two examples of lattices for the two-dimensional case are shown in \cref{fig:lattices}.

\subsubsection{Two-dimensional quantum-spin Hall}
\begin{figure}
    \centering
    \includegraphics[width=0.97\linewidth]{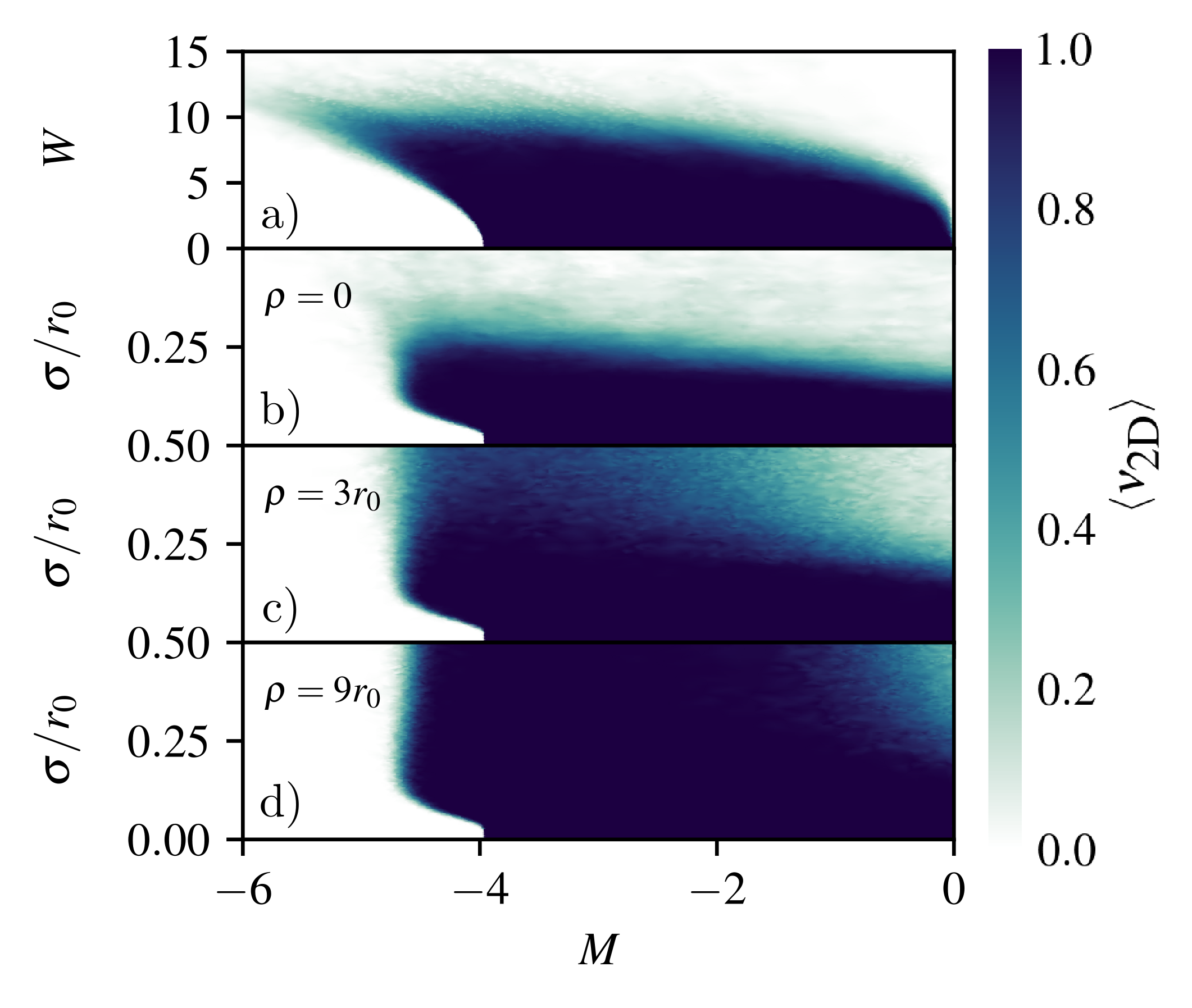}
    \caption{
    Phase diagram of the localizer index for the two-dimensional BHZ model comparing the effect of on-site Anderson and structural disorder, with a small fraction of spin-frame disordered sites $\eta=0.15$, for a system size $L=30$. Each point is calculated with $A=1$, and averaged over 100 realisations of disorder. The phase diagram is calculated in each case as a function of the parameter $M$ in \cref{eqn:ham_onsite} and the respective type of disorder. The spectral localizer index given by \cref{eq:index_loc} is computed for $\kappa=1$ and $E, \mathbf{x}$ at the centre of the energy spectrum and the system, respectively.
    (a) The localizer index for the BHZ model as a function of Anderson disorder strength $W$.
    (b-d) The localizer index as a function of $M$ and the strength of Gaussian structural disorder, characterized by its standard deviation $\sigma$ with no inter-site repulsion $\rho$. Three cases are shown with  $\rho = 0$, $3r_0$ and $9r_0$.
    Note that in the thermodynamic limit, the phase boundary in the case of no disorder would sit at exactly at $M=-4$, however it is slightly rescaled in finite size systems.
    }
    \label{fig:phasediagramBHZ}
\end{figure}
In the case of Anderson disorder, shown in \cref{fig:phasediagramBHZ} (a), we find a familiar story \cite{li_topological_2009,groth_theory_2009,Prodan2011,Huang2018,Huang2018b,yamakage_z2_2012,Setescak2025}. Here, provided that we start with $M$ approximately $\in [-5.4, -4]$, the system can be driven into a topological phase by increasing the Anderson disorder coefficient $W$. 

In contrast, in the case of structural disorder we find strikingly different results depending on the type of structural disorder implemented. 
With pure Gaussian disorder ($\rho=0$), shown in \cref{fig:phasediagramBHZ} (b) 
we find that the system initially has an expanded topological regime, indicating a structural analogue of the topological Anderson insulator transition. 
This expanded regime may be understood by considering how the connectivity of the graph changes with structural disorder. On the square lattice all vertices connect to four neighbours. However as we start to introduce structural disorder, the connectivity initially increases. This can be seen in \cref{fig:lattices}, where the number of edges increases with small amounts of disorder, as well as in detail in Appendix \ref{apx:lattices}. Increased connectivity has the effect of rescaling the coupling strength, where each site feels a stronger coupling on average. Since the topological transition is driven by the competition between coupling terms (parametrised by $A$) and on-site terms (parametrised by $M$), increased connectivity can be viewed as effectively diminishing $M$ with respect to $A$, enlarging the topological regime in which strong hopping dominates.

As Gaussian disorder ($\sigma$) is further increased, the system at $\rho = 0$ is driven into a topologically trivial phase. However, when the inter-site repulsion $\rho$ is increased, allowing the vertices to effectively avoid one another as they are disordered, we find that Gaussian structural disorder is increasingly unable to drive the system into a trivial phase. This is shown in \cref{fig:phasediagramBHZ} (c-d). This effect suggests that the transition to a trivial phase in \cref{fig:phasediagramBHZ} (b) originates from the tendency of truly random pointsets to form highly internally-connected clusters, which are then relatively weakly connected to one another, effectively breaking the bulk insulator up into islands. When sites repel one another in the disordering procedure, this effect is supressed, the material remains uniformly connected, and the topological phase survives. This is particularly relevant, since by all measures, the disorder generated for large $\rho$ is closer to that found in covalently bonded amorphous materials~\cite{zallen_physics_1998}, where atoms are prohibited from getting arbitrarily close to one another. In turn, metallic glasses and systems close to the crystallization temperature are expected to be less ordered at the atomic scale, and hence we expect them to behave more similarly to the case $\rho\approx 0$.

\subsubsection{Three-dimensional \texorpdfstring{$\mathbb{Z}_2$}{Z2} topological insulator}

\begin{figure}
    \centering
    \includegraphics[width=0.97\linewidth]{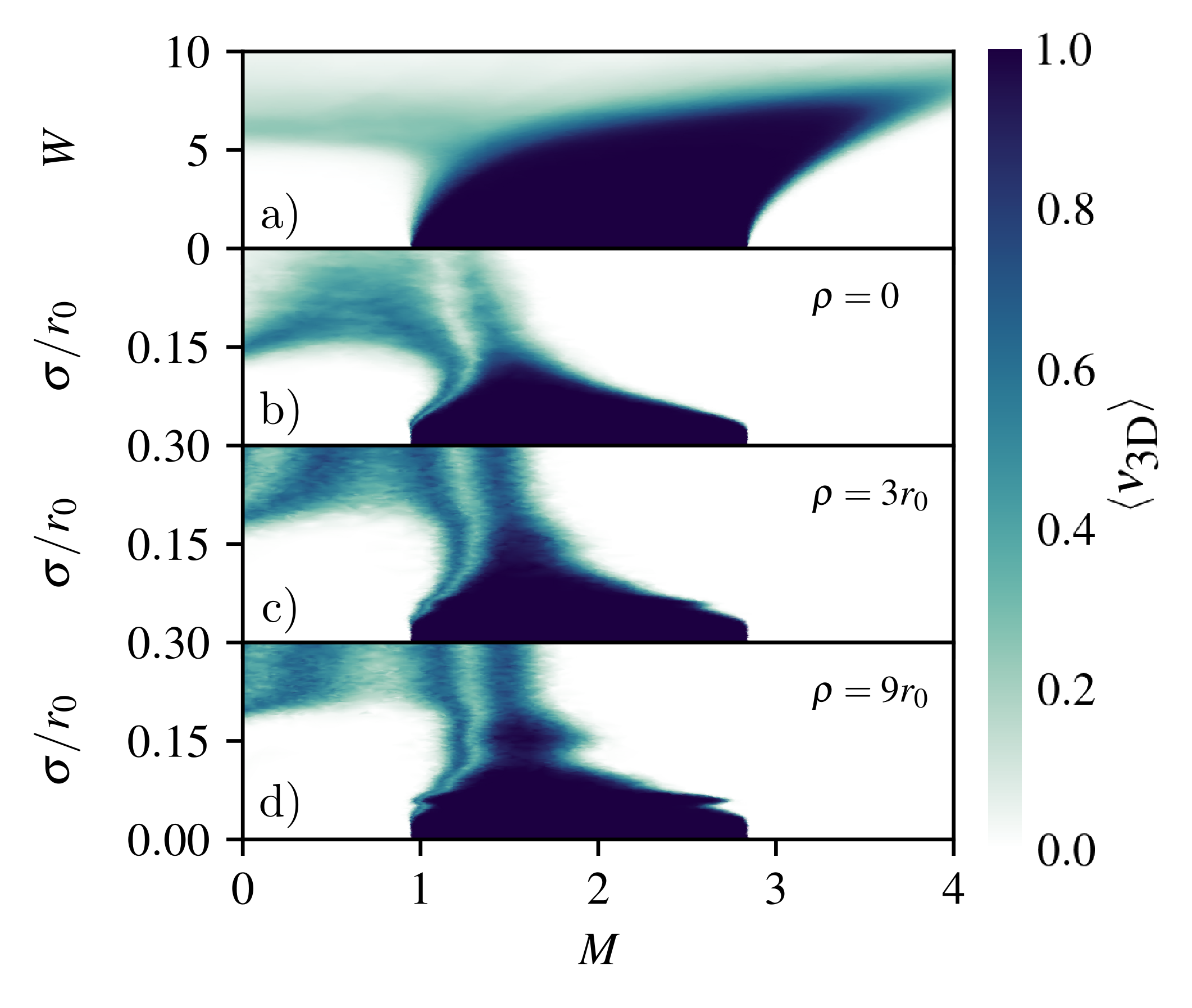}
    \caption{
        Phase diagram of the localizer index for the three-dimensional topological insulating model in the presence of Anderson and structural disorder. Each point is calculated with $A=1$, and averaged over 100 realizations of disorder, for a system size $L=10$.  The localizer index, \cref{eq:signdet}, is computed for $\kappa=2$ and $E, \mathbf{x}$ at the bulk of the energy spectrum and the system, respectively.
        (a) Localizer index for the three-dimensional topological insulator model as a function of Anderson disorder strength $W$.
        (b-d) Localizer index as a function of $M$ and the strength of Gaussian structural disorder, characterized by its standard deviation $\sigma$ with the inter-site repulsion $\rho$. As in \cref{fig:phasediagramBHZ}, three cases are shown with  $\rho = 0$, $3r_0$ and $9r_0$. 
        As in \cref{fig:phasediagramBHZ}, phase boundaries are slightly shifted by a combination of finite-size effects and the rescaling of bond strength by the distance as in \cref{eqn:coupling_strength}.
    }
    \label{fig:phasediagram3D}
\end{figure}

In the three-dimensional case with on-site disorder, the phase diagram obtained from the localizer, shown in \cref{fig:phasediagram3D} (a), is in good agreement with previous studies of disordered three-dimensional $\mathbb{Z}_2$ topological insulators~\cite{Guo2010,Kobayashi2013, Akagi2017,Setescak2025}. 
Similar to the two-dimensional case, we find that within  a range of values of the mass parameter, $ 4 \gtrsim M \gtrsim 3$, the system can start in a trivial phase and be driven into a topological phase by increasing on-site disorder \cite{Guo2010,Kobayashi2013, Akagi2017,Setescak2025}.

The effect of structural disorder is strikingly different to the two-dimensional BHZ model. First, the phase diagram with structural disorder, see \cref{fig:phasediagram3D}(b-d) differs from the phase diagram obtained with on-site disorder case, \cref{fig:phasediagram3D}(a). Specifically, it is no longer possible to drive the system into the topological Anderson insulator at $ 4 \gtrsim M \gtrsim 3$.
Second, we observe that the results obtained with purely Gaussian disorder ($\rho=0$) and with strong inter-site repulsion ($\rho=3r_0$ and $9r_0$) in \cref{fig:phasediagram3D}(b-d) are qualitatively very similar. In all cases, increasing structural disorder eventually drives the system from the topological phase into a trivial phase. 

These results suggest that the effect of $\rho$ is not as efficient in reinstating the topological phase in three dimensions as it is in two-dimensions.
This happens even thought increasing $\rho$ is highly effective at ensuring that states remain evenly spaced apart, as we show  in Appendix \ref{apx:lattices}.
Hence, three-dimensional topological insulators see amorphicity as truly detrimental, regardless of the ways in which the system is structurally disordered. 
An alternative route towards inducing a topological phase in three-dimensions is to increase $r_0$ in a random point-set, as discussed in Ref.~\cite{mukati_topological_2020}.
However, when $r_0$ is increased substantially the phase space for the topological phase seems to be reduced, similar to our findings.

\section{Discussion and Conclusions \label{sec:conclusions}}

We have discussed how different kinds of structural disorder affect the topological phase diagrams of disordered non-crystalline systems, and how local-frame disorder can compromise the tools we use to diagnose these phases. Specifically, we found that in two dimensions increasing the local bond-connectivity acts to increase the region of phase space dedicated to a topological phase. This effect likely occurs because increasing connectivity boosts the energy scales corresponding to spin-orbit-coupling hopping terms, which then can more easily win over on-site terms that work to trivialise the system. We have found that this connectivity effect is absent in three-dimensional systems. 

In our calculations we have included the scrambling of the local spin degree of freedom. We have shown that this novel type of disorder, introduced in Ref.~\cite{Schirmann2025} and expected to be realistic in amorphous materials, artificially compromises the diagnosis of topological phases based on spin-projection methods. While this is a known issue, and some workarounds exist~\cite{Bau2024b}, we showed that the spectral localizer is immune to such scrambling. The spectral localizer directly incorporates the time-reversal operator in its definition, rather than relying on a spin-projection~\cite{LORING2015,Cerjan_2024_Tutorial,wong2026}, and can be computed numerically efficiently~\cite{wong2026}. This property allowed us to derive a consistent definition for a $\mathbb{Z}_2$ time-reversal invariant phase at any strength of spin-frame disorder, or any other term that mixes spin species in the material.

We have refrained from exhaustively analysing the precise phase diagram boundaries and also the possible differences between the transitions driven by Anderson on-site disorder and structural disorder. For Chern insulators in class A early numerics suggested that critical exponents may differ from those of the quantum Hall transition~\cite{ivaki_criticality_2020}, although more in-depth studies seem to align with conventional expectation~\cite{Bera2024}. In clarifying these questions, it would be informative for future work to develop similar studies for the quantum-spin Hall insulators in class AII studied here, aided by the spectral localizer.

To conclude, we have found that not all structural disorder is born equal. Structural disorder that penalises site-clustering can help to realize non-crystalline two-dimensional quantum spin-Hall insulators in solid-state and synthetic platforms. This type of structural disorder is expected in more covalently bonded materials~\cite{zallen_physics_1998}, suggesting a design principle for material realizations. Thus, our results suggest that disordered topological materials with well-defined local order may be much more robust compared to 
completely random structural disorder, which could be expected in solids that are metallic glasses or closer to crystallization temperatures.
Furthermore, we have found that as a material is disordered, the bond connectivity often increases, useful as a guiding experimental principle in future attempts to realize non-crystalline topological phases in two-dimensions. 

Alongside these results comes a caution, however, that we find that there is no `universal consequence' of structural disorder. Depending on the implementation of disorder, and the dimensionality of the system, we have found that it is possible to realise a wide variety of physical effects. Some forms of structural disorder destroy topology, some do not. Thus, when considering a specific material realisation, our results emphasise the importance of tailoring the type of structural disorder used in modelling to the microscopic properties of the material in question.

\section{Acknowledgments}
We thank A. Cerjan and P. Wulles for inspiring discussions about the applicability of topological markers, as well as I. Araya Day and A. Akhmerov for assistance on numerical methods.
All authors are supported by the European Research Council (ERC) Consolidator grant under grant agreement No. 101042707 (TOPOMORPH). \\

\textit{Code availability} -- The codes and data to reproduce the figures in this work can be found in Ref.~\cite{leveraging_zenodo2026}. Numerical calculations were performed using the python packages Kwant~\cite{groth_kwant_2014} and Adaptive~\cite{Nijholt2019}.

\bibliography{BHZ_loc.bib}
%

\clearpage
\newpage

\appendix
\setcounter{secnumdepth}{5}
\renewcommand{\theparagraph}{\bf \thesubsubsection.\arabic{paragraph}}

\setcounter{figure}{0}
\renewcommand{\thefigure}{S\arabic{figure}}
\renewcommand{\theHfigure}{S\arabic{figure}}

\section{Generating Amorphous systems} \label{apx:lattices}

\begin{figure}
    \centering
    \includegraphics[width=\linewidth]{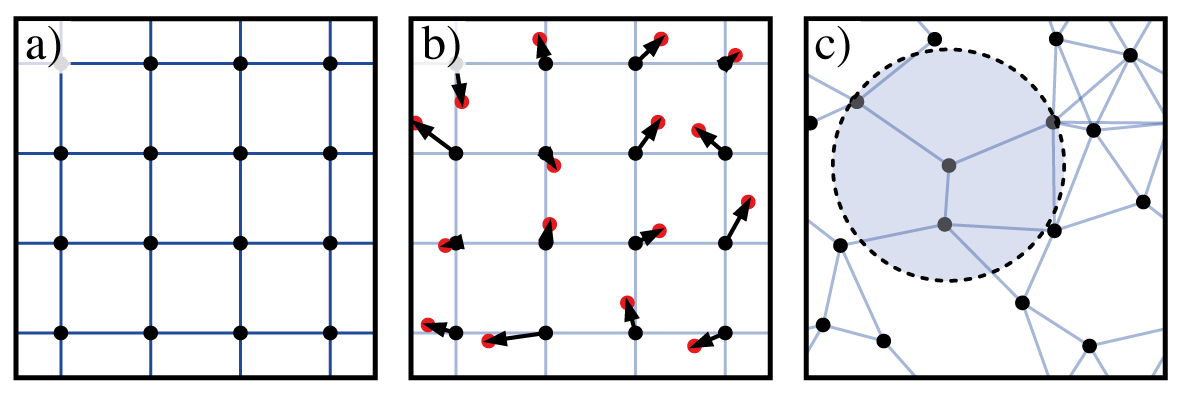}
    \caption{Disordering a square lattice. (a) The initial square lattice. 
    (b) Vertices are moved to a new set of positions, chosen randomly.
    (c) New bonds are created, where every pair of vertices closer than $1.3r_0$, with $r_0$ the original lattice spacing, are connected with a bond. The distance cut-off is indicated with the dashed circle. 
    }
    \label{fig:apx_lattice_1}
\end{figure}

To transform an ordered square or cubic lattice gradually into a disordered one, we perform an operation in three steps. First, we discard the edges, and only consider the positions of the vertices. Next we introduce a random shift to the positions of each vertex in the system, producing a disordered array of points. Finally, we construct edges between every pair of vertices that is closer than a threshold value---in practice we use $1.3\, r_0$, where $r_0$ is the lattice spacing of the original square lattice. This is depicted in \cref{fig:apx_lattice_1}.

The only non-trivial part of this procedure is the random shift applied to each vertex, which is performed vertex-by-vertex, iterating over every site in the point-set. A full iteration involves applying a random shift to every vertex in a random order. The probability distribution used is a Gaussian distribution centered on the original position, however we include a modification that disfavours placing a vertex too close to any other vertices. Additionally, this is done by applying a small shift several times, ensuring that the system is able to find a random configuration independent of the order in which vertices were shifted.

Thus, let us explain the procedure for generating the probability distribution that is used to sample the new position of each vertex in the algorithm. 

\begin{figure}
    \centering
    \includegraphics[width=\linewidth]{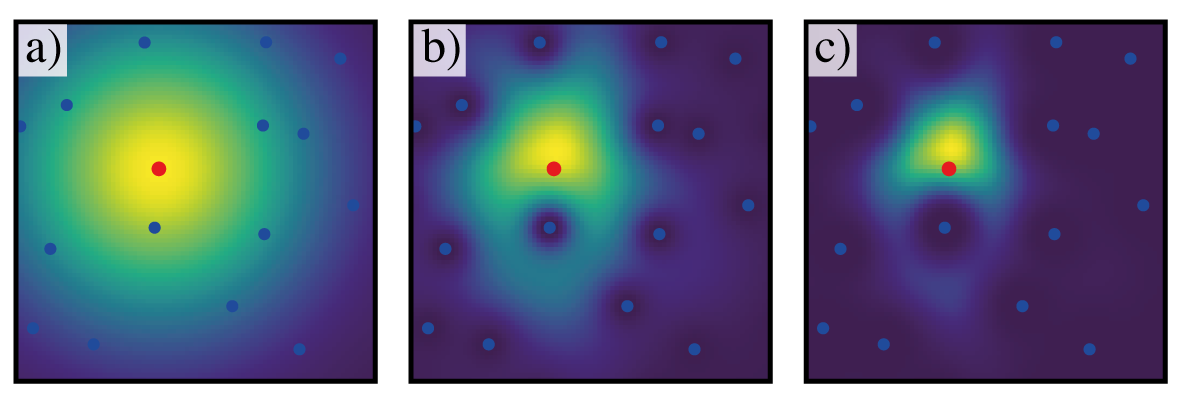}
    \caption{The probability distribution generated for moving a single site (marked red) to a new position on the unit square, $\textbf x^{\textup{new}}$. In all three cases, $\sigma = 0.3$.
    (a) With $\rho = 0$, the probability distribution is a Gaussian around the original position of the site.
    (b) With $\rho=0.1$, the probability is highly suppressed around a distance approximately $\sim 0.1$ around each vertex.
    (c) For large $\rho=0.3$, the probability distribution is strongly shaped by the potential produced by the vertices in the system.
    }
    \label{fig:apx_probs}
\end{figure}

In order to tune between the square lattice and a uniform distribution of positions we use a Gaussian distribution. Here, starting with a vertex at position $\textbf x^0$, we draw a probability distribution given by,
\begin{align}
    p(\textbf x^{\textup{new}} ) = \frac{1}{\mathcal N} e^{
    -
    {\left |
    \textbf {x}^{\textup{new}}
    - 
    \textbf {x}^{0} 
    \right | ^2}/
    {2\sigma ^2}
    },
\end{align}
where $\sigma$ parametrises the standard deviation of the Gaussian, and $\mathcal N$ is a normalisation factor. In analogy with statistical mechanics, we may rewrite this  in the form of a partition function (with $\beta = 1$) of a classical harmonic oscillator (HO),
\begin{align}
    p(\textbf x^{\textup{new}} ) = \frac{1}{\mathcal N}e^{-\mathcal H_\textup{HO}},
\end{align}
with a `Hamiltonian' given by
\begin{align}
    \mathcal H_{\textup{HO}} = -\frac
    {\left |
    \textbf {x}^{\textup{new}}
    - 
    \textbf {x}^{0} 
    \right | ^2}
    {2\sigma ^2}.
\end{align}
Now let us modify this potential to prohibit $\textbf x^{\textup{new}}$ from being too close to the other vertices in the system, which sit at positions $\textbf x^j$, with $j\in 1...N$. Thus, we introduce a second term to the Hamiltonian, which adds an energy penalty in the regions close to each $\textbf x^j$. We consider an inter-site repulsive interaction,
\begin{align}
    \mathcal H \rightarrow \mathcal H_{\textup{HO}} + H_{\textup{Int}},
\end{align}
with 
\begin{align}
    H_{\textup{Int}} = \frac{S_d}{2\pi}\sum_{j} \frac{\rho}{\left |
    \textbf {x}^{\textup{new}}
    - 
    \textbf {x}^{j} 
    \right |^{d-1}},
\end{align}
where $\rho$ is a parameter that controls the strength of this repulsion, $d$ is the dimensionality of space ($d \in \{2,3\}$ in this work), and $S_d$ is the surface area of a sphere in $d$ dimensions. Thus, we arrive at a new probability distribution of the form
\begin{align}
    p(\textbf x^{\textup{new}} ) = \frac{1}{\mathcal N}e^{-\mathcal H_\textup{HO} - \mathcal H_{\textup{Int}}}.
\end{align}
The resulting probability distribution for three values of $\rho$ is shown in \cref{fig:apx_probs}. 

This shift is applied $n_{\mathrm{it}}$ times, where at each point we find a new set of vertices $\textbf x^{\textup{new}}$, and then repeat the same process. Depending on the value of $n_{\mathrm{it}}$, we must also rescale the standard deviation of the shift, to ensure that repeated translations correspond roughly to a single step with standard deviation $\sigma$, thus we use a $\sigma_{\mathrm{eff}} = \sigma / \sqrt{n_{\mathrm{it}}}$, and similarly rescale $\rho_{\mathrm{eff}} = \rho / n_{\mathrm{it}}$.

\begin{figure}
    \centering
    \includegraphics[width=\linewidth]{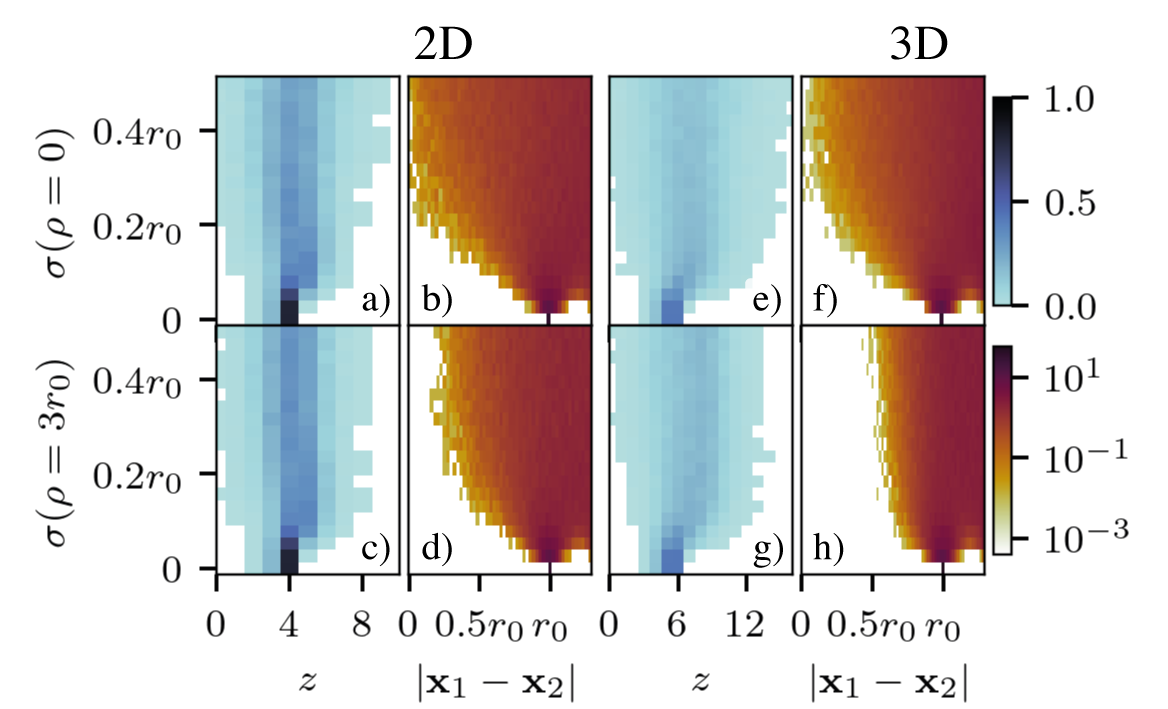}
    \caption{The probability distribution of connectivity (in blue) and bond distance (in green) as a function of $\sigma$. For bond distances we plot the logarithm of the probability. For each case we average over 10 example graphs. In two dimensions (a-d) we start with a $20\times20$ grid of points. In the three-dimensional case (e-h) we start with a $10 \times10\times10$ grid. Two cases are shown for each dimensionality, that of no inter-vertex repulsion (top row) and the case with the repulsion set to $\rho = 3r_0$ (bottom row).
    }
    \label{fig:connectivity}
\end{figure}

The effect of increasing $\rho$ is illustrated in \cref{fig:connectivity}, where we plot the distribution of connectivity and bond lengths in two and three dimensions respectively. We see that the effect on the average connectivity is relatively similar with and without repulsion $\rho$, where at small disorder the average quickly moves up from 4 (6 in three dimensions) to around 5.5 (8 in three dimensions). The most drastic effect is seen in the bond distances, where we find that in the case of $\rho = 0$, where we find that around $\sigma \sim 0.25 r_0$, the bond distances can reach arbitrarily small values. However when repulsion is included we find that at large disorder, there is a minimum distance between vertices that is not crossed.

\begin{figure}
    \centering
    \includegraphics[width=\linewidth]{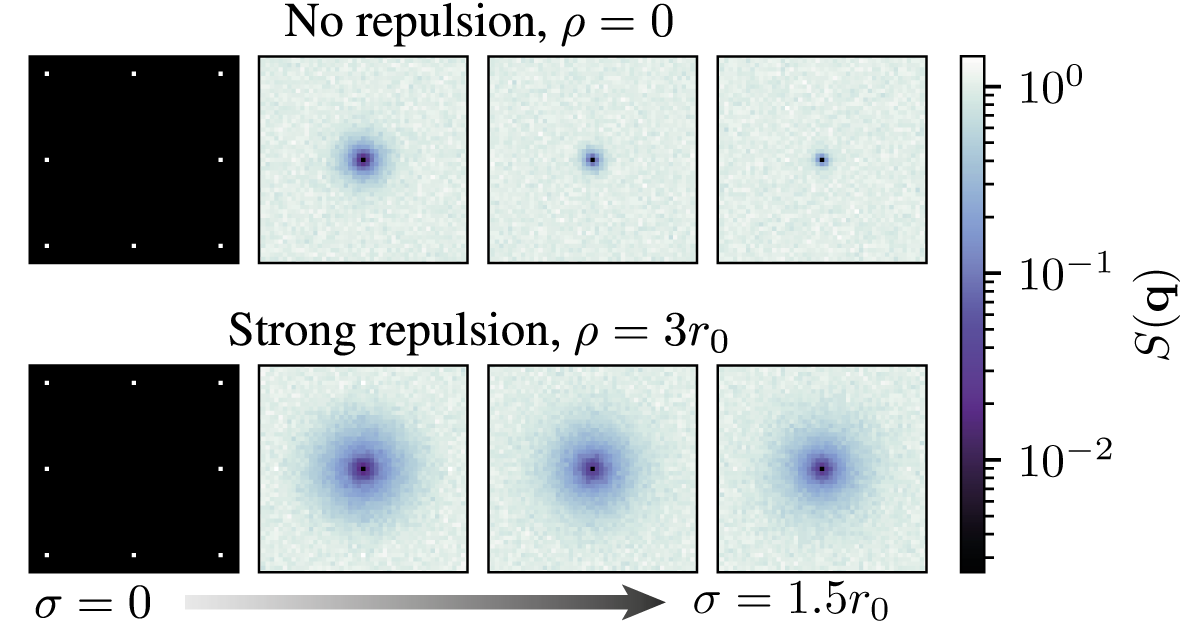}
    \caption{Structure factor for an ensemble of point-sets that are progressively disordered starting from a uniform square lattice. With no inter-site repulsion, $\rho = 0$, the distribution is initially hyperuniform, indicated by the region around the origin where the structure factor tends towards zero. As disorder is increased this hyper uniformity disappears as the system becomes closer to a set of uniformly sampled points. When sites repel one another, $\rho = 3r_0$, we find that the system remains hyperuniform up to arbitrarily large disorder.
    }
    \label{fig:structure_factors}
\end{figure}

The effect on the structure factor is shown in \cref{fig:structure_factors}, where we plot the structure factor for a set of point-sets generated starting from a uniform grid of size $20 \times 20$ and progressively disordering the positions of the vertices---the real-space example is shown in \cref{fig:lattices}. In the case of $\rho = 0$, we see that the initial square lattice distribution is initially broken to a distribution whose structure factor falls to zero around the origin, indicating a hyperuniform distribution. Here, the fluctuations in point density vanish at large length scales. 
However, as we increase the standard deviation of the disordering translations, $\sigma$, this hyperuniformity shrinks, indicating that the distribution becomes closer and closer to a true uniform distribution. On the other hand, when we disorder the square lattice with a strong inter-site repulsion $\rho = 3r_0$, we form a hyperuniform distribution that remains for arbitrarily large $\sigma$. 

\section{Choosing \texorpdfstring{$\kappa$}{k} for disordered systems}
\label{app:kappa}
 
Constructing the spectral localizer requires choosing an appropriate value of the parameter \texorpdfstring{$\kappa$}{k}. As mentioned in the main text, this coefficient is responsible for weighting the Hamiltonian and position operator in the spectral localizer's spectrum. From \cref{eq:general_localizer}, we can see that for \texorpdfstring{$\kappa$}{k} $\rightarrow 0$, the spectrum of $\mathcal L$ simply reduces to that of the Hamiltonian. Instead, if \texorpdfstring{$\kappa$}{k} is too large, the spectral localizer only contains information of the system's sites. Hence, the size of $\kappa$ must reflect a balance of the typical energy and length scales of the system.

Refs.~\cite{Setescak2025, schulz2024topological, loring2020spectral, Cerjan_2024_Tutorial} derived bounds for \texorpdfstring{$\kappa$}{k}. Combined with numerical benchmarks~\cite{Cerjan_2024_Tutorial} these point out that the expression
\begin{align}
    \label{eqn:optimal_kappa}
    \kappa \approx\frac{E_{gap}}{L},
\end{align}
is typically an optimal choice, where $E_{gap}$ and $L$ corresponds to the band gap of the Hamiltonian and the system's size respectively. A relatively large window of \texorpdfstring{$\kappa$}{k} values around the condition in \cref{eqn:optimal_kappa} typically reproduces well the crystalline topological phase diagram.

The $\kappa$ range of suitable values can be visualized through the computation of the local gap, the smallest eigenvalue of the spectral localizer in \cref{eqn:2D_localizer}. The local gap serves as a quantitative measure of the robustness of the topological phase identified by the spectral localizer~\cite{Cerjan_2024_Tutorial}. Intuitively, the larger the local gap, the more resistant the spectral localizer is to a gap closing---a prerequisite for a phase transition. 

Here, we find the optimal choice of \texorpdfstring{$\kappa$}{k} by approximately matching the localizer gap closings with the known phase transitions of the crystalline models we study. We note that a single choice of $\kappa$ necessarily introduces an ambiguity in the position of the phase boundaries. This ambiguity is inherent to real-space methods to calculate invariants in finite systems. For example, the local Chern marker~\cite{bianco11} requires summing over a chosen bulk real-space area. As we approach the phase transition the edge states delocalize into the bulk, and the chosen area becomes polluted with edge state weight; different area choices result in different phase boundaries. Similarly, we now show how different $\kappa$ result in slightly different phase boundaries. 
%

\begin{figure}
    \centering
    \includegraphics[width=\linewidth]{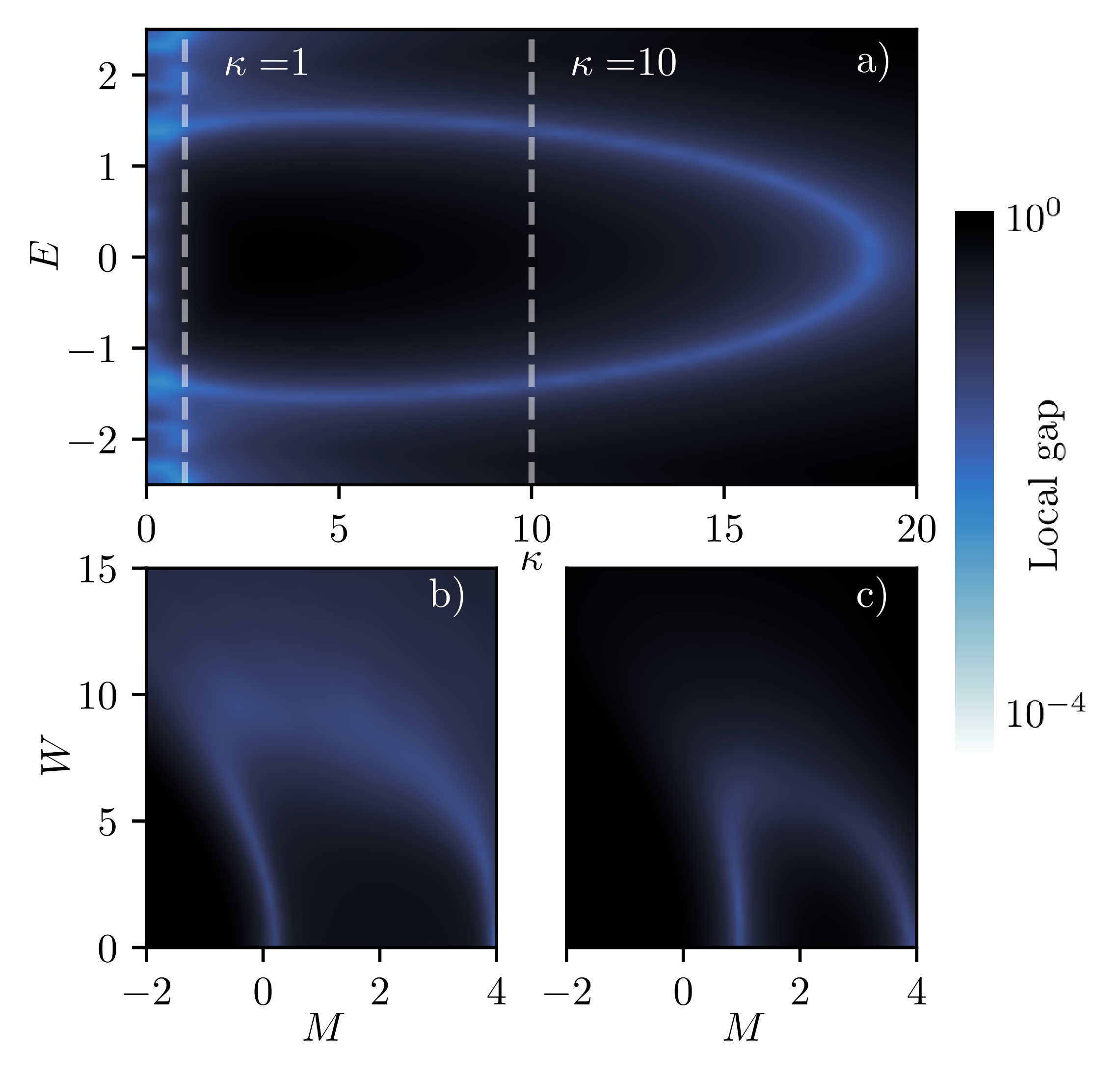}
    \caption{
    (a) Gap of the localizer from \cref{eqn:2D_localizer} as a function of \texorpdfstring{$\kappa$}{k} and energy shift $E$ for the crystalline BHZ model, with $A=M=1$, in logarithmic scale.
    (b) Local gap as a function of the onsite disorder strength $W$ and mass $M$ for $\kappa=1$.
    (c) Local gap as a function of the onsite disorder strength $W$ and mass $M$ for $\kappa=10$.
    The spectral localizer gap is computed for $\mathbf{x}$ at the center of the system and averaged over 100 realizations.
    } 
    \label{fig:fig_kappa}
\end{figure}

In \cref{fig:fig_kappa}a, we plot the local gap as a function of $\kappa$ and energy $E$ for the Bernevig-Hughes-Zhang (BHZ) model in a topological phase. A distinct bubble structure emerges, separating the topological region from the trivial one. Any value of $\kappa$ within this bubble correctly diagnoses the topological phase and yields a large local gap, making it a valid choice for computing the spectral localizer.
To identify the most appropriate value, we further evaluate the local gap as a function of the onsite disorder $W$ and mass $M$ for two representative values, $\kappa=1$ and $\kappa=10$, as shown in \cref{fig:fig_kappa}b-c. By comparing these results with the phase diagram in \cite{yamakage_z2_2012}, we find that $\kappa=1$ provides the best agreement, so that we choose this value for our calculations. 

The choice of $\kappa$ remains valid even in the presence of disorder, where $E_{gap}$ is renormalized, since it is effectively constrained by the local gap, which stays finite up to the topological transition. At the boundaries of the topological phase, however, this choice becomes more important, as the mobility gap closes and the spectral localizer approaches a singular configuration. As with finite-size effects, discussed above, disorder will also decrease the ability of local markers to correctly diagnose the phase close to phase boundaries.

\begin{figure}
    \centering
    \includegraphics[width=\linewidth]{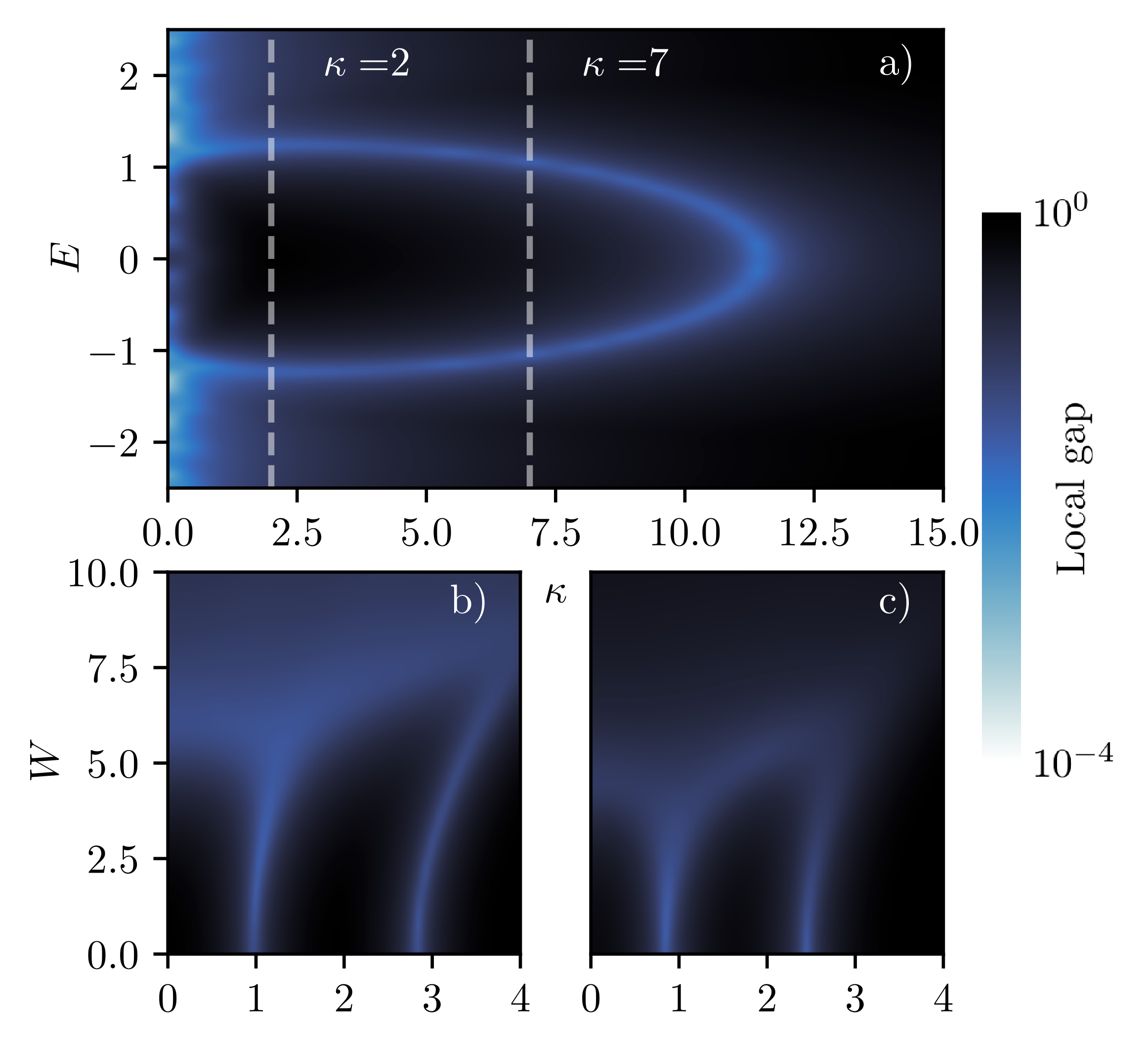}
    \caption{
    (a) Gap of the localizer from \cref{eq:3D_localizer} as a function of \texorpdfstring{$\kappa$}{k} and energy shift $E$ for the crystalline three-dimensional topological insulator, with $A=1$, $M=1.5$, in logarithmic scale.
    (b) Local gap as a function of the onsite disorder strength $W$ and mass $M$ for $\kappa=2$.
    (c) Local gap as a function of the onsite disorder strength $W$ and mass $M$ for  $\kappa=7$.
    The spectral localizer gap is computed for $\mathbf{x}$ at the center of the system and averaged over 100 realizations.
    } 
    \label{fig:fig_kappa_3D}
\end{figure}

The same procedure is applied to the three-dimensional model in \cref{fig:fig_kappa_3D}, where we find that $\kappa=2$ yields the phase diagram most consistent with the known topological phase boundaries in the absence of disorder.

\section{\label{app:locsubtelties}The Spectral Localizer Under Arbitrary Unitary Rotations}
A subtlety emerges when calculating the spectral localizer for a system in class AII with added local on-site rotations. In order to understand this subtlety, we must first consider in detail the motivation behind \cref{eqn:rot_localizer_main}. Once the background is established, we shall derive the formalism for calculating the spectral localiser in a rotated basis.

\subsection{A Recap of the Spectral Localizer for Class AII} \label{sec:localiser_recap}

We start by restating the expression for the spectral localizer in two dimensions \cite{cerjan2023spectral,LORING2015}, \cref{eqn:2D_localizer} in the main text, 
\begin{align}
    \mathcal L &= \begin{pmatrix}
        H & X -iY\\
        X +iY & -H\\
    \end{pmatrix}, \\
    & = \sigma_x \otimes X +  \sigma_y \otimes Y +  \sigma_z \otimes H
\end{align}
where we have absorbed the energy shift $E$ and $x$ and $y$ shifts into the matrices $H$, $X$ and $Y$. The index is given by the sign of the Pfaffian of $\mathcal L$. However, before computing the Pfaffian, we must first rotate $\mathcal L$ into an antisymmetric form, such that the Pfaffian is well-defined \cite{LORING2015}.

Let us consider the symmetries of our system. Firstly, $H$ has spinful time-reversal symmetry, $\hat {\mathcal T} = w \mathcal K$, where $\mathcal K$ is complex conjugation, and $w$ is a unitary matrix, conventionally $w = is^y$. For the system to be in class AII, the time reversal operator must satisfy,
\begin{align}
    \hat {\mathcal T}^2 = -\1 \iff w^T = -w. \label{eqn:trs_w}
\end{align}
Thus, the Hamiltonian must satisfy,
\begin{align}
    [\hat {\mathcal T}, H] = 0 \iff w^\dag H w = H^*,
\end{align}
where $^*$ denotes complex conjugation without transposition. 

This symmetry is directly inherited by the spectral localizer $\mathcal L$, where the time-reversal symmetry of the Hamiltonian leads to a particle-hole symmetry (PHS) in the localizer. To construct the operator for PHS, consider
\begin{align}
    \hat {\mathcal C} &=v\mathcal K,\\
    \textup{with } v & = \sigma_y \otimes w,
\end{align}
which satisfies
\begin{align}
    \{
    \mathcal L
    ,
    \hat{\mathcal C} 
    \} &= 0, \label{eqn:chiral_sym}\\
    \textup{with } \hat{\mathcal C}^2 & = + \1.
\end{align}
Furthermore, the localizer has neither chiral symmetry $\hat {\mathcal S} $, nor time reversal symmetry. To see why, note that they have to come together since $\hat {\mathcal S}  =\hat {\mathcal C} \hat {\mathcal T}$, the localizer construction spans all three Pauli matrices, so cannot have sub-lattice symmetry (unless $H$ started with it).
Thus, we see that the system being in class AII means that the localizer is in class D, with PHS that squares to $+1$. 

A useful property of a matrix in class D is that it can always be rotated to a basis in which it is pure imaginary and antisymmetric---a necessary condition to compute a Pfaffian. In this basis, the rotated localizer will anti-commute with the complex conjugation operator $\mathcal K$. Thus, let us propose that there exists a special unitary matrix $Q$, which rotates us to a `canonical basis' in which the localizer is pure imaginary and antisymmetric,
\begin{align}
    \mathcal L_C &= Q^\dag \mathcal L Q,\label{rotated_loc_1}\\
     \textup{such that } \{\mathcal L_C , \mathcal K \}&=0\label{rotated_loc_2}.
\end{align}
Note that we require $Q$ to have determinant 1, since the sign of the Pfaffian can be arbitrarily changed by rotation with a unitary matrix according to $\pf{QXQ^\dag} = \det(Q) \pf X$. Inserting \cref{rotated_loc_1} into \cref{rotated_loc_2}, we arrive at the following condition for $Q$,
\begin{align}
    \{ Q^\dag \mathcal L Q , \mathcal K \} &= 0 \\
    \implies \{\mathcal L , Q\mathcal KQ^\dag \} &= 0. 
\end{align}
Thus, comparing to \cref{eqn:chiral_sym}, for $\mathcal L_C$ to satisfy \cref{rotated_loc_2} we must have $Q\mathcal KQ^\dag$ equal to our original $\hat {\mathcal C}$ operator. Thus the condition on our rotation $Q$ is that it must satisfy
\begin{align}\label{eqn:qqtv}
    QQ^T = v.
\end{align}
To progress, note that $v = v^T$, since $v = \sigma_y \otimes w$, and both $\sigma_y$ and $w$ are antisymmetric, see \cref{eqn:trs_w}. Thus, our desired $Q$  corresponds to a Takagi factorisation of the unitary symmetric $v$, which is always possible \cite{takagi_algebraic_1924,ikramov_takagi_2012}. We provide an explicit algorithm for constructing $Q$ in the next section. 

Thus, we have the guarantee that $\mathcal L_C = Q^\dag \mathcal L Q$ is pure imaginary and anti-symmetric, allowing us to calculate the spectral localizer index as, 

\begin{align}
(-1)^{\nu} = \text{sign}\big(\pf{i\mathcal L_C}\big) \in \pm 1
\end{align}

\subsubsection{Explicitly Calculating the Takagi Factorisation of \texorpdfstring{$v$}{v}}

The matrix $v$ satisfies two properties, unitarity ($vv^\dag = \1$) and symmetry ($v = v^T \implies v^* = v^\dag$). Thus, we start by showing that $v$ admits a real eigenbasis. Consider a general eigenvector of $v$, with
\begin{align}
    v\ket{\alpha} = e^{i\alpha} \ket{\alpha}.
\end{align}
Taking the conjugate and applying our two conditions we find that,
\begin{align}
    v^*\ket{\alpha}^* & = e^{-i\alpha} \ket{\alpha}^*,\\
    \implies
    e^{i\alpha}\ket{\alpha}^*  &=  v\ket{\alpha}^.
\end{align}
Thus, we find the conjugated eigenstate also has eigenvalue $e^{i\alpha}$. We have two possibilities: Either $\ket \alpha \in \mathbb R$, or $\ket \alpha\in \mathbb C$. In the case that $\ket \alpha$ is complex, then its conjugate is a degenerate eigenstate, and so we can construct a new pair of eigenstates,
\begin{align}
    \ket{\beta} = \frac{1}{\sqrt 2} (\ket \alpha + \ket \alpha^*), \\
    \ket{\gamma} = \frac{-i}{\sqrt 2} (\ket \alpha - \ket \alpha^*).
\end{align}
Either way, we find that it is always possible to diagonalise $v$ in a real basis. Therefore, we can construct a real matrix $S \in SO(n)$, which diagonalises $v$,
\begin{align}
    v = R e^{i\theta}R^T,
\end{align}
Next, we take the unitary square root of $v$, arriving at the correct transformation,
\begin{align}
    Q = R e^{i\frac{\theta}{2}} R^T,
\end{align}
which satisfies \cref{eqn:qqtv}, and so must satisfy our conditions \cref{rotated_loc_1,rotated_loc_2}.

\subsubsection{The Case of Conventional Time Reversal Symmetry}%
As a check, let us compare the results of this section with the pre-existing methods proposed in \cite{LORING2015}. For most time-reversal-symmetric-invariant systems, the form of $w$ is known, 
\begin{align}
    w = is_y.
\end{align}
This means that the PHS operator obeyed by the spectral localizer is given by 
\begin{align}
    v = i\sigma_y\otimes s_y.
\end{align}
This can be easily diagonalised in terms of the eigenvalues and eigenvectors of $\sigma_y$, which has $\lambda = \pm 1$, with eigenvectors $\ket{\pm_y}$. Given this, the eigenvalues of $v$ are $\begin{pmatrix}
    i&i&-i&-i
\end{pmatrix}$, and the rotation matrix $R$ is given by the eigenvalues formed by the four possible outer products of two copies of $\ket{\pm_y}$, where we may take advantage of the degeneracy to find a real matrix,
\begin{align}
    R = \begin{pmatrix}
        1&&1& \\
        &1&&1 \\
        1&&-1& \\
        &1&&-1 \\
    \end{pmatrix}
\end{align}
Now, it is straightforward to calculate $Q$, which is given by
\begin{align}
    Q = R\, \textup{diag}\begin{pmatrix}
        e^{i\frac \pi 4}&e^{i\frac \pi 4}
        &e^{-i\frac \pi 4}&e^{-i\frac \pi 4}
    \end{pmatrix}R^T,
\end{align}
which is evaluated as
\begin{align}
    Q = \begin{pmatrix}
        1&&&-i \\
        &1&i& \\
        &i&1& \\
        -i&&&1 \\
    \end{pmatrix}\label{eqn:conventional_q}
\end{align}
This recovers the exact form of $Q$ given in \cite{LORING2015}. This operator acts on the combined localiser and spin basis, and so can equivalently be expressed in the form,
\begin{align}
    Q = \1 + i \sigma^y \otimes s^y.
\end{align}

\subsection{Time Reversal Under a Basis Change}

In general, the definition of the time reversal symmetry operator is chosen to be 
\begin{align}
    \hat{\mathcal T} = i\sigma_y \mathcal K,
\end{align}
acting on the spin basis. However, this is a basis-dependent construction, so it is worth considering what happens when we subject the Hamiltonian to an arbitrary unitary transformation,
\begin{align}
    H \rightarrow H' = U HU^\dag.
\end{align}
Considering the effect of the time-reversal operator, we find that
\begin{align}\begin{aligned}\relax
    [ \hat{\mathcal T}, H ] = [\hat{\mathcal T}, U^\dag H'U] &= 0
    \\ 
    \implies [U\hat{\mathcal T} U^\dag,  H'] &= 0
\end{aligned}
\end{align}
Therefore, we see that the time-reversal operator is modified in this new basis to 
\begin{align}
    \hat {\mathcal{T}} \rightarrow \hat {\mathcal{T}}' = U i\sigma_y U^T \mathcal K.
\end{align}
Therefore, we find that the form of the time-reversal operator remains unchanged only if the chosen unitary satisfies
\begin{align}
    U \sigma_y U^T = \sigma_y.
\end{align}

\subsection{The Localizer Under a Basis Change}

Let us now consider the general case, of a system with some arbitrary choice of time-reversal symmetry operator, $\hat {\mathcal T} = w \mathcal K$, which undergoes some arbitrary unitary transformation $U$. Furthermore, we will make the assumption that $U$ acts only onsite, 
\begin{align}
    U = \sum_{\textbf r} \ketbra{\textbf r}{\textbf r}\otimes U_{\textbf r}, 
\end{align}
where $\textbf r$ denotes site location. This ensures that $[U,X] = [U,Y] = 0$. Thus, the localizer is also transformed under this change of basis,
\begin{align}
    \mathcal L \rightarrow \mathcal L' = (\sigma_0 \otimes U) \mathcal L (\sigma_0 \otimes U^\dag).
\end{align}
Let us follow the effect of this transformation step-by-step through the derivation presented in \cref{sec:localiser_recap}.
As before, we find that the time reversal operator $w$ is transformed to 
\begin{align}
    w \rightarrow w' = UwU^T,
\end{align}
and so the particle-hole operator obeyed by the full spectral localizer is also modified to
\begin{align}
    v \rightarrow v' &= \sigma_y \otimes UwU^T, \\ 
    &= (\sigma_0 \otimes U) v (\sigma_0 \otimes U^T).
\end{align}
Thus, we can calculate the change to the diagonalisation of $v$, where the rotation matrix $R$ becomes,
\begin{align}
    R \rightarrow R' = UR,
\end{align}
and so the final $Q$ operator is modified to 
\begin{align} \label{eqn:q_transformation}
    Q \rightarrow Q' = UR e^{i\frac{\theta}{2}} R^T U^T.
\end{align}
As before, this operator is guaranteed to bring the spectral localizer to canonical form,
\begin{align}
    Q'^\dag \mathcal L' Q' &= \mathcal L_C',\\
    \textup{ with } \{\mathcal L_C', \mathcal K\} &= 0.
\end{align}

\subsubsection{A Note on the Hadamard Gate}
In the paper, we primarily consider the effect of a Hadamard gate, as a simple example of a local unitary transformation,
\begin{align}
    \mathsf H = \frac{1}{\sqrt 2}\begin{pmatrix}
        1 & 1 \\ 
        1 & -1
    \end{pmatrix}.
\end{align}
The first thing we need to calculate is the effect of $\mathsf H$ on the time-reversal symmetry operator, given by
\begin{align}
    \mathsf H i s_y \mathsf H = -is_y.
\end{align}
Thus, we see that if the Hadamard gate is applied to all sites in the system, the form of the time-reversal symmetry operator changes only by a global sign, and we may continue to use the conventional form of $Q$ given in \cref{eqn:conventional_q}, since global phases have no effect here. However, let us consider applying a Hadamard to a subset of sites in the system, 
\begin{align}
    U = \sum_{\textbf r} \ketbra {\textbf r}{\textbf r} \otimes \left [ 
    a_{\textbf r} \mathsf H + (1-a_{\textbf r}) \1
    \right ],
\end{align}
where $a_{\textbf r}$ is a random variable $\in \{0,1\}$ that decides which sites undergo a Hadamard transformation. If we start with the canonical time-reversal symmetry operator, $i\sigma_y$, we find that the modified time-reversal symmetry operator will be
\begin{align}
    w = \sum_{\textbf r} \ketbra {\textbf r}{\textbf r} \otimes (-1)^{a_{\textbf r}} i\sigma_y,
\end{align}
and so $Q$ must be transformed according to \cref{eqn:q_transformation}. 

\section{The Chern Number, Bott Index and Chern marker for Class A} \label{app:markers}

In this Appendix we provide a short background on the Bott index and local Chern marker \cite{Kitaev20062,bianco11,loring2019guidebottindexlocalizer}, showing how they may be derived from the Chern number and providing a useful numerical trick for simplifying the calculation of the Chern marker in periodic boundaries. 

\subsection{The Bott Index Calculates the Chern Number}

We start by stating the conventional formalism for the Chern number of a single band in a two-dimensional translationally symmetric topological insulator (Class A). Let us make the assumption that we are working in a system with $L\times L$ unit cells, and periodic boundaries---the extension to non-square geometry is straightforward. Translational symmetry allows us to decompose the eigenstates forming our chosen band into a combination of a plane wave and unit cell part,
\begin{align}
    \ket{\psi_{\textbf k}} = \ket {\textbf k} \otimes \ket{u_{\textbf k}},
\end{align}
where the unit cell component $\ket{u_{\textbf k}}$ contains all topological information, and the plane wave component is defined as 
\begin{align}
    \ket {\textbf k} = \frac{1}L\sum_{\textbf m} e^{-i \textbf m \cdot \textbf k} \ket {\textbf m}.
\end{align}
Here, $\textbf m = \begin{pmatrix}
    m_x, m_y
\end{pmatrix} \in \mathbb Z^2$ indexes the unit cell, and allowed $\textbf k$ values are quantised to \begin{align}
    \textbf k= \frac{2\pi}{L} \begin{pmatrix}
    n_x, n_y
\end{pmatrix}.
\end{align}

The Chern number of this band is given by the sum of the Berry curvature across the full Brillouin zone 
\cite{asboth_short-course_2016},
\begin{align}\label{eqn:chern_number}
    \mathcal C = \frac{1}{2\pi} \sum_{\textbf k} \gamma^B_{\textbf k},
\end{align}
with the berry phase $\gamma_B$ defined according to 
\begin{align}
    \begin{aligned}\label{eqn:berry_phase}  
        r_{\textbf k} e^{i\gamma^B_{\textbf k}} = &\braket{u_{\textbf k}}{u_{\textbf k+\bm \delta_x}}
        \braket{u_{\textbf k+\bm \delta_x}}{u_{\textbf k+\bm \delta_x +\bm \delta_y}} \\ 
        &\cdot 
        \braket{u_{\textbf k+\bm \delta_x +\bm \delta_y}}{u_{\textbf k+\bm \delta_y}}
        \braket{u_{\textbf k+\bm \delta_y}}{u_{\textbf k}} ,
    \end{aligned}
\end{align}
where $r$ is some real component and the vector $\bm \delta_{\mu}$ is the shortest possible displacement in momentum space in either the $k_x$ or $k_y$ directions,
\begin{align}
    \bm \delta_\mu = \delta \hat {\textbf e}_{\mu},
\end{align}
with $\delta = \frac {2\pi}{L} $ and  $\hat {\textbf e}_{\mu}$ being a unit vector in the $x$ or $y$ direction. To construct the Bott index, we must find a real-space expression for \cref{eqn:chern_number}. Thus, let us consider the effect of the two operators, $e^{i\delta X}$ and $e^{i\delta Y}$, which one can show take the following form in momentum space,
\begin{align}
    e^{i\delta X} = \sum_{\textbf k}\ketbra{\textbf k + \bm \delta_x}{\textbf k}, \\
    e^{i\delta Y} = \sum_{\textbf k}\ketbra{\textbf k + \bm \delta_y}{\textbf k}. 
\end{align}

Let us now consider the full projector onto our band
\begin{align}
    P = \sum_{\textbf k} \ketbra {\textbf k}{\textbf k} \otimes u_{\textbf k},
\end{align}
where we have defined $u_{\textbf k} = \ketbra{u_{\textbf k}}{u_{\textbf k}}$ in order to simplify the following notation. By considering the effect of our momentum space translation operators, we can show the following identity,
\begin{align} \label{eqn:p_shift}
    e^{-i\delta X} P e^{i\delta X} = \sum_{\textbf k} \ketbra {\textbf k}{\textbf k} \otimes u_{\textbf k + \bm \delta_{x}},
\end{align}
with a similar result for $e^{i\delta Y}$. 

Now we may construct the Bott index by considering an operator of the form
\begin{align}
    UVU^\dag V^\dag = P e^{-i\delta X} P e^{-i\delta Y} Pe^{i\delta X} P e^{i\delta Y} P.
\end{align}
Using \cref{eqn:p_shift}, we show that this operator takes the following form in momentum space,
\begin{align}
    UVU^\dag V^\dag = \sum_{\textbf k}\ketbra{\textbf k}{\textbf{k}}\otimes
    u_{\textbf k }
    u_{\textbf k +\bm \delta_{x}}
    u_{\textbf k +\bm \delta_{x} + \bm \delta_{y}}
    u_{\textbf k +\bm \delta_{y}}
    u_{\textbf k }.
\end{align}
Next, recalling the definition of the Berry phase, \cref{eqn:berry_phase}, we find that this operator may be written as
\begin{align}\label{eqn:UVUV}
     UVU^\dag V^\dag
     = \sum_{\textbf k}\ketbra{\textbf k}{\textbf{k}}\otimes u_{\textbf k} \cdot r_{\textbf k} e^{i\gamma^B_{\textbf k}}.
\end{align}
Thus, we see that each diagonal element of the operator on the right hand side contains exactly the Berry curvature for the corresponding point in momentum space. All that remains is to extract only the sum of Berry curvatures, which is done in two steps. First we take the logarithm in order to extract only the Berry phase component. The left hand side of \cref{eqn:UVUV} has zeros on the support of all unoccupied states, so a logarithm of this matrix will diverge. Thus, we add the projector $Q = \1-P$ before taking the logarithm, ensuring the $Q$ component does not contribute,
\begin{align}
    \log \left (
    UVU^\dag V^\dag + Q
    \right ) = \sum_{\textbf k}\ketbra{\psi_{\textbf k}}{\psi_{\textbf k}}  \left ( \log r_{\textbf k} + i\gamma^B_{\textbf k}\right ).
\end{align}
Finally, we take the imaginary part of the trace, extracting only the sum of the Berry phase over the full Brillouin zone,
\begin{align}
    \mathcal B &= \frac{L^2}{2\pi}\im \Tr \log \left [ UVU^\dag V^\dag + Q\right],\\
    \implies \mathcal B &= \frac{1}{2\pi} \sum_{\textbf k} \gamma^B_{\textbf k} = \mathcal C
loo\end{align}
Thus, we have shown that in the case of a translationally symmetric system with periodic boundaries, the Bott index evaluated at any point in our material always exactly calculates the Chern number.

\subsection{Equivalence of the Bott Index to the Chern Marker} \label{sec:bott_chern_equivalence}

The Bott index is numerically extremely expensive to compute for large matrices as a consequence of the matrix logarithm. Thus, a useful approximation can be made in the limit of $L\gg 1 \implies \delta \ll 1$ . Consider the following three matrices,
\begin{align}
    P_{\textbf x} &= e^{-i\delta X} Pe^{i\delta X},\\ 
    P_{\textbf {xy}} &= e^{-i\delta (X+Y)} Pe^{i\delta (X+Y)},\\
    P_{\textbf {y}} &= e^{-i\delta Y} Pe^{i\delta Y}.
\end{align}
The Bott index can be conveniently rewritten in terms of these three operators as
\begin{align}
    \mathcal B = \frac{L^2}{2\pi}\im \Tr \log \left [ P P_{\textbf x}P_{\textbf {xy}} P_{\textbf {y}} P+ Q\right]. \label{eqn:b_restated}
\end{align}
Next, we can Taylor expand the operator inside the logarithm in $\delta$, using the identities
\begin{align}
    P_{\textbf x} &=  P-i\delta [X,P] + \mathcal O(\delta^2),\\ 
    P_{\textbf {xy}} &= P-i\delta [X+Y,P] + \mathcal O(\delta^2),\\
    P_{\textbf {y}} &= P-i\delta [Y,P] + \mathcal O(\delta^2).
\end{align}
Thus, we can expand the term inside the logarithm in \cref{eqn:b_restated} up to second order as,
\begin{align}
\begin{aligned}
    P P_{\textbf x}P_{\textbf {xy}} P_{\textbf {y}} P+ Q
    = \1 &- 2\delta^2 P [X,P] [Y,P] P  \\ 
        &+ \mathcal O(\delta^3) + (\textup{Hermitian terms})
\end{aligned}
\end{align}
where we have used the identity $P[O,P]P = 0$ for any operator $O$, and neglected to write the Hermitian terms which vanish when the imaginary part is taken. Finally, inserting this into \cref{eqn:b_restated} and using the identity $\log (\1 + x) \approx x$, we may express the Bott index in the form,
\begin{align}
    \mathcal B \approx \mathcal C =  2 \pi i \Tr \left [P [X,P] [Y,P] P - h.c. \right ],
\end{align}
the exact form of the Chern marker \cite{bianco11}.
\vspace{0.4cm}

\begin{figure}
    \includegraphics[width=\columnwidth]{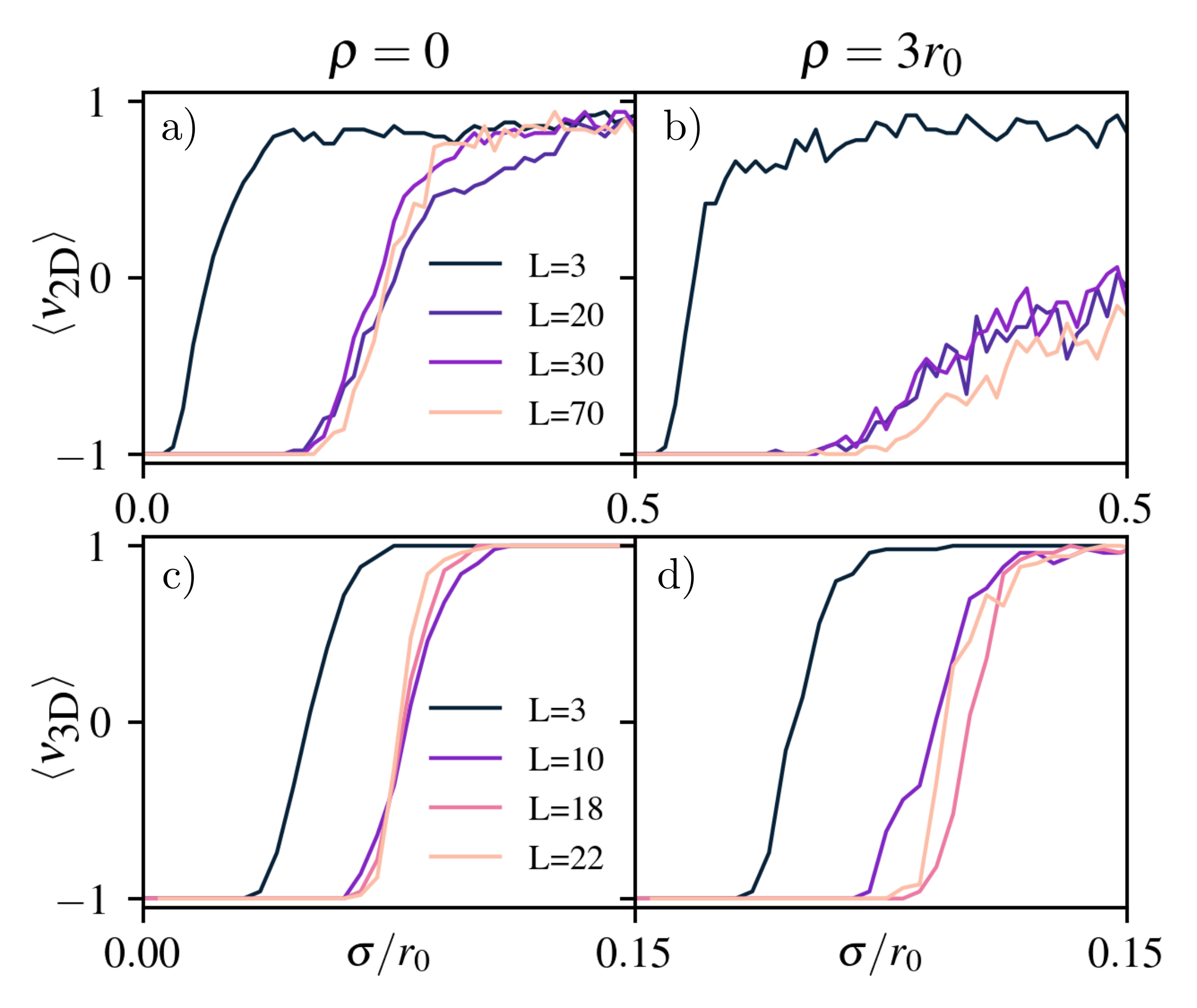}
    \caption{
     Finite size scaling of the spectral localizer invariant for the BHZ model, with $M=-2$, as a function of the strength of Gaussian structural disorder, characterized by its standard deviation $\sigma$ with (a) no inter-vertex repulsion $\rho=0$ and b) strong inter-vertex repulsion $\rho=3r_0$.
     Finite size scaling of the spectral localizer invariant for the three-dimensional model, with $M=2$, as a function of the Gaussian structural disorder strength $\sigma$ with (a) no inter-vertex repulsion $\rho=0$ and b) strong inter-vertex repulsion $\rho=3r_0$.
     The spectral localizer is computed for $E, \mathbf{x}$ at the centre of the energy spectrum and the system, respectively. The scaling parameter $\kappa$ takes the value 1 for the BHZ model while for the three-dimensional topological insulator is 2. 
    } 
    \label{fig:fig_finitesize}
\end{figure}%

\subsubsection*{Periodic Boundaries}

The above quantity is valid in an infinite system, where $X$ and $Y$ are well-defined, however a subtlety emerges when working in periodic boundaries, since both position operators are discontinuous across the boundary where they jump from $L$ back to $0$. In order to address this, we can make a small modification to our expression. Noticing that the commutators are effectively `dressing' the projector with a displacement,
\begin{align}
    [X, P] = \sum_{jk} (x_j - x_k)P_{jk} \ketbra{\textbf r_j}{\textbf r_k}.
\end{align}
If $x_j$ and $x_k$ are on opposite sides of the discontinuity where real space was `stitched' in periodic boundaries, then the displacement $(x_j - x_k)$ is effectively incorrect. In order to rectify this, we may define the periodic $\bm \Delta_{jk}$, defined as the shortest displacement from $\textbf r_j$ to $\textbf r_k$, taking into account the periodic boundaries. Assuming the periodic boundaries are defined on $[0,L)$ in each axis, we can define $ \bm \Delta_{jk}$ by simply folding all displacements back into the interval $[-L/2, L/2)$
\begin{align}
    \bm \Delta_{jk}^x + \frac L2 = x_j - x_k + \frac L2 \mod L,
\end{align}
with a similar definition for $\bm \Delta_{jk}^y$.
The Chern marker must then be restated using the following \textit{periodic commutators},
\begin{align}
    [X,P]_{\textup{PB}} = \sum_{jk} \Delta_{jk}^x P_{jk} \ketbra{\textbf r_j}{\textbf r_k},\\
    [Y,P]_{\textup{PB}} = \sum_{jk} \Delta_{jk}^y P_{jk} \ketbra{\textbf r_j}{\textbf r_k}.
\end{align}
Thus, we arrive at a restatement of the Chern marker which is valid in periodic boundaries,

\begin{align}
    \mathcal C_{\textup{PB}} =  2 \pi i \Tr \left ( P [X,P]_{\textup{PB}} [Y,P]_{\textup{PB}} \right ) + h.c.
\end{align}

\section{\label{apx:finitesizescaling} Finite-size scaling in the presence of structural disorder} 

To ensure that the phase diagrams presented in the main text are not influenced by finite-size effects, we perform a finite-size scaling analysis of the spectral localizer invariant for both the two- and three-dimensional models in \cref{fig:fig_finitesize}. The scaling is evaluated as a function of structural disorder along a one-dimensional cut of the phase diagrams shown in \cref{fig:phasediagramBHZ} and \cref{fig:phasediagram3D}, with the mass parameter held fixed at the midpoints of the phase boundaries. We do not consider onsite disorder here, as its effects have already been studied in detail~\cite{li_topological_2009,groth_theory_2009,Guo2010,Prodan2011,yamakage_z2_2012,Huang2018,Huang2018b}.

For both models, we find that the invariant converges rapidly with $L$, with finite-size effects only observed for very small system sizes ($L \approx 3$). This robustness is expected to be weaker at the boundaries of the phase, where the edge states are less localized and the system's gap is smaller than at the midpoint.

In addition, $\kappa$ is kept fixed for all system sizes considered. The observed consistency across system sizes demonstrates that this choice is robust and well-controlled, in agreement with the discussion in Appendix \ref{app:kappa}, further validating its use throughout the main text.

\end{document}